\documentclass[useAMS,usenatbib]{mn2e}
\usepackage[final]{graphics}
\usepackage{epsfig}

\newcommand{\he}[1] {He\,{\sc #1}}
\newcommand{\hel}[2] {He\,{\sc #1}~$\lambda$#2}
\newcommand{\kms}{\mbox{$\mathrm{km~s^{-1}}$}}
\newcommand{\Porb}{\mbox{$P_{\rm orb}$}}
\newcommand{\Line}[3]{\Ion{#1}{#2}~$\lambda$#3}

\newcommand{\Ion}[2] {#1\,{\sc #2}}
\newcommand{\Ha}{\mbox{${\mathrm H\alpha}$}}
\newcommand{\Hb}{\mbox{${\mathrm H\beta}$}}
\newcommand{\Hg}{\mbox{${\mathrm H\gamma}$}}

\newcounter{fistro}
\title[The SW Sextantis stars]{SW\,Sextantis stars: the dominant population of CVs with orbital periods between 3--4\,hours}
\author[P. Rodr\'\i guez-Gil et al.]{P. Rodr\'\i guez-Gil$^{1,2}$\thanks{E-mail:prguez@iac.es},
B. T. G\"ansicke$^{2}$,
H.-J. Hagen$^{3}$,
S. Araujo-Betancor$^{1}$,
\newauthor
A. Aungwerojwit$^{2,4}$,
C. Allende Prieto$^{5}$,
D. Boyd$^{6}$,
J. Casares$^{1}$,
D. Engels$^{3}$,
\newauthor
O. Giannakis$^{7}$,
E. T. Harlaftis$^{8}$,
J. Kube$^{9}$,
H. Lehto$^{10,11}$,
I. G. Mart\'\i nez-Pais$^{1,12}$,
\newauthor
R. Schwarz$^{13}$,
W. Skidmore$^{14}$, A. Staude$^{13}$ and M. A. P. Torres$^{15}$\\
$^{1}$Instituto de Astrof\'\i sica de Canarias, V\'\i a L\'actea s/n, La Laguna, E-38205, Santa Cruz de Tenerife, Spain\\
$^{2}$Department of Physics, University of Warwick, Coventry CV4 7AL, UK\\
$^{3}$Hamburger Sternwarte, Universit\"at Hamburg, Gojenbergsweg 112, 21029 Hamburg, Germany\\
$^{4}$Department of Physics, Faculty of Science, Naresuan University, Phitsanulok, 65000, Thailand\\
$^{5}$McDonald Observatory and Department of Astronomy, University of Texas, Austin, TX 78712, USA\\
$^{6}$British Astronomical Association, Variable Star Section, West Challow, OX12 9TX, England, UK\\
$^{7}$Institute of Astronomy and Astrophysics, National Observatory of Athens, P.O. Box 20048, Athens 11810, Greece\\
$^{8}$Institute of Space Applications and Remote Sensing, National Observatory of Athens, PO Box 20048, Athens 11810, Greece\\
$^{9}$Alfred-Wegener-Institut f\"ur Polar- und Meeresforschung, B\"urgermeister-Smidt-Stra{\ss}e 20, 27568 Bremerhaven, Germany\\
$^{10}$Tuorla Observatory, University of Turku, FIN-21500 Piikki\"o, Finland\\
$^{11}$Department of Physics, FIN-20014 University of Turku, Finland\\
$^{12}$Departamento de Astrof\'\i sica, Universidad de La Laguna, Tenerife, Spain\\
$^{13}$Astrophysikalisches Institut Potsdam, An der Sternwarte 16, 14482 Potsdam, Germany\\
$^{14}$California Institute of Technology, Mail Code 105-24, Pasadena, CA 91125-24, USA\\
$^{15}$Harvard-Smithsonian Center for Astrophysics, 60 Garden St, Cambridge, MA 02138, USA}
\begin{document}
\date{Accepted 2007. Received 2007}
\pagerange{} \pubyear{2007}
\maketitle
\begin{abstract}
We present time-series optical photometry of five new CVs identified by the Hamburg Quasar Survey. The deep eclipses observed in HS 0129+2933 (= TT Tri), HS 0220+0603, and HS 0455+8315 provided very accurate orbital periods of 3.35129827(65), 3.58098501(34), and 3.56937674(26) h, respectively. HS 0805+3822 shows grazing eclipses and has a likely orbital period of 3.2169(2) h. Time-resolved optical spectroscopy of the new CVs (with the exception of HS 0805+3822) is also presented. Radial velocity studies of the Balmer emission lines provided an orbital period of 3.55 h for HS 1813+6122, which allowed us to identify the observed photometric signal at 3.39 h as a negative superhump wave. The spectroscopic behaviour exhibited by all the systems clearly identifies them as new SW Sextantis stars. HS 0220+0603 shows unusual N\,{\sc ii} and Si\,{\sc ii} emission lines suggesting that the donor star may have experienced nuclear evolution via the CNO cycle.

These five new additions to the class increase the number of known SW Sex stars to 35. Almost 40 per cent of the total SW Sex population do not show eclipses, invalidating the requirement of eclipses as a defining characteristic of the class and the models based on a high orbital inclination geometry alone. On the other hand, as more SW Sex stars are identified, the predominance of orbital periods in the narrow 3--4.5 h range is becoming more pronounced. In fact, almost {\it half} the CVs which populate the 3--4.5 h period interval are definite members of the class. The dominance of SW Sex stars is even stronger 
in the 2--3 h period gap, where they make up 55 per cent of all known gap CVs. These statistics are confirmed by our results from the Hamburg Quasar Survey CVs. Remarkably, 54 per cent of the Hamburg nova-like variables have been identified as SW Sex stars with orbital periods in the 3--4.5 h range. The observation of this pile-up of systems close to the upper boundary of the period gap is difficult to reconcile with the standard theory of CV evolution, as the SW Sex stars are believed to have the highest mass transfer rates among CVs.

Finally, we review the full range of common properties that the SW Sex stars exhibit. Only a comprehensive study of this rich phenomenology will prompt to a full understanding of the phenomenon and its impact on the evolution of CVs and the accretion processes in compact binaries in general.   
\vspace{6cm}
\end{abstract}

\begin{keywords}
accretion, accretion discs -- binaries: close -- stars: individual: HS 0129+2933, HS 0220+0603, HS 0455+8315, HS 0805+3822, HS 1813+6122 -- novae, cataclysmic variables
\end{keywords}

\section{Introduction}

\begin{figure*}
\includegraphics[width=5.8cm]{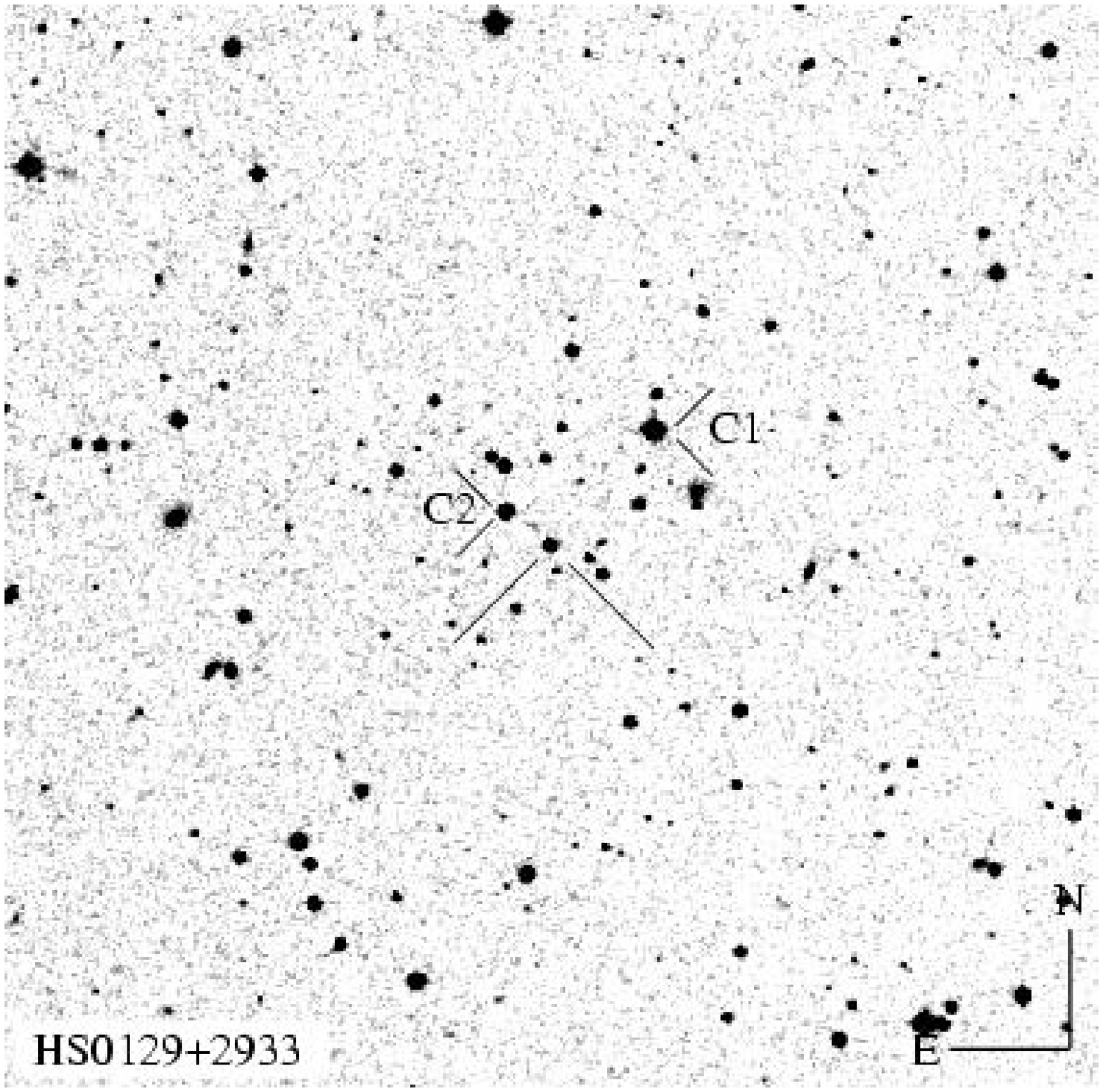}
\hfill
\includegraphics[width=5.8cm]{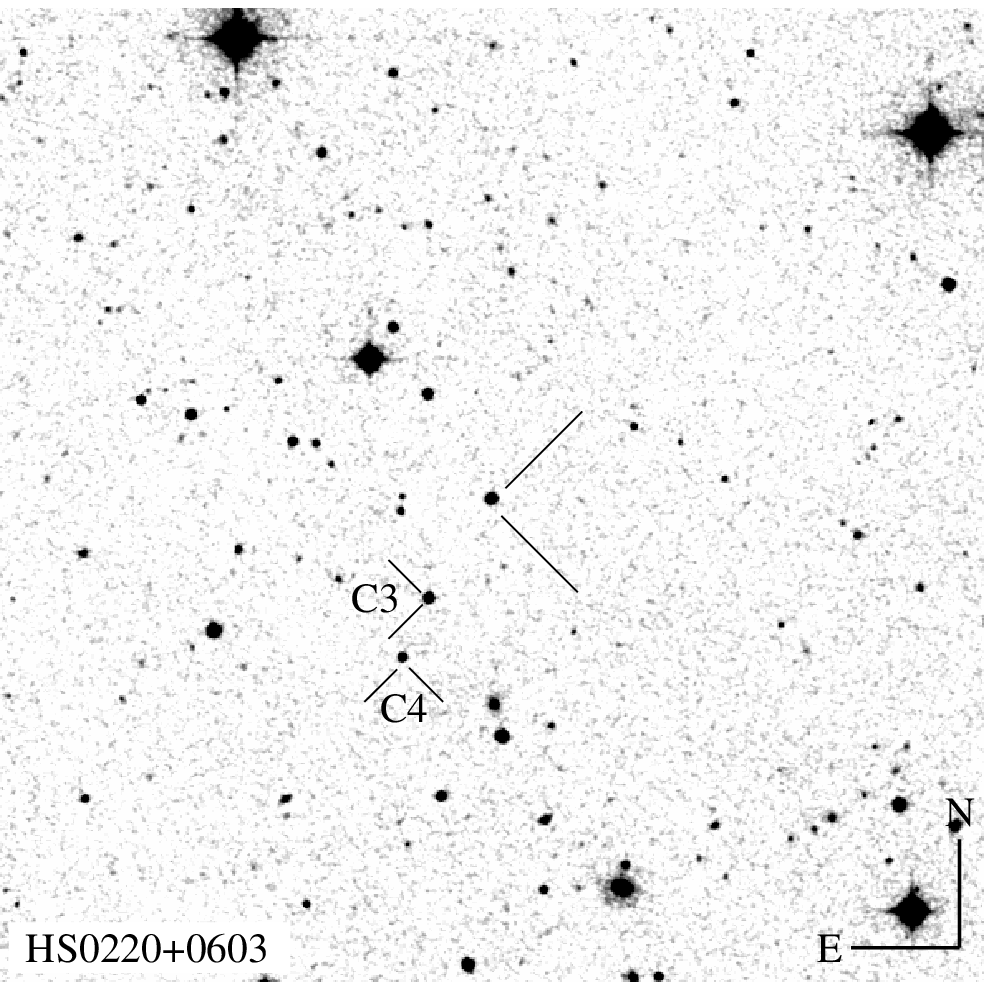}
\hfill
\includegraphics[width=5.8cm]{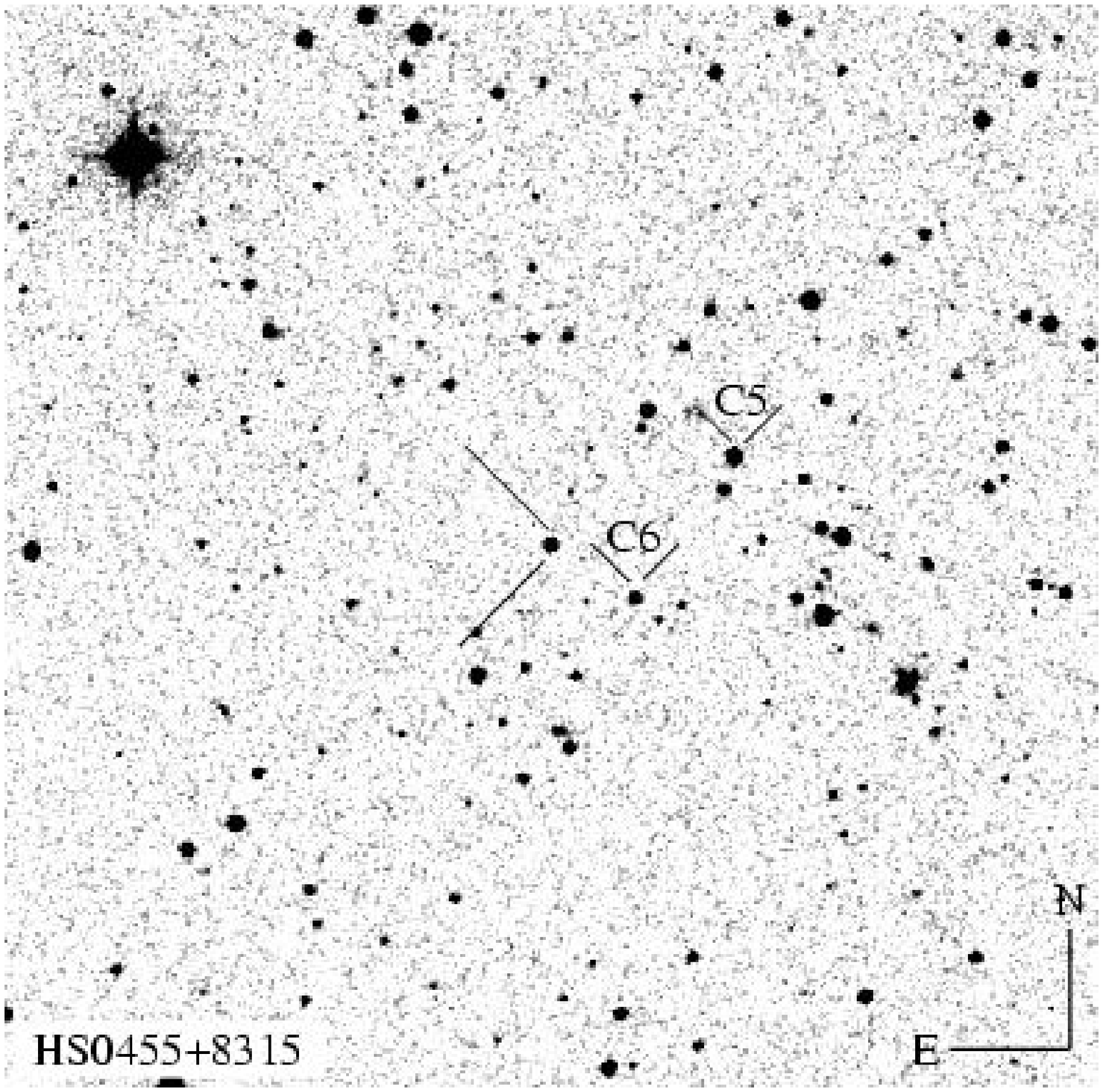}
\hfill
\includegraphics[width=5.8cm]{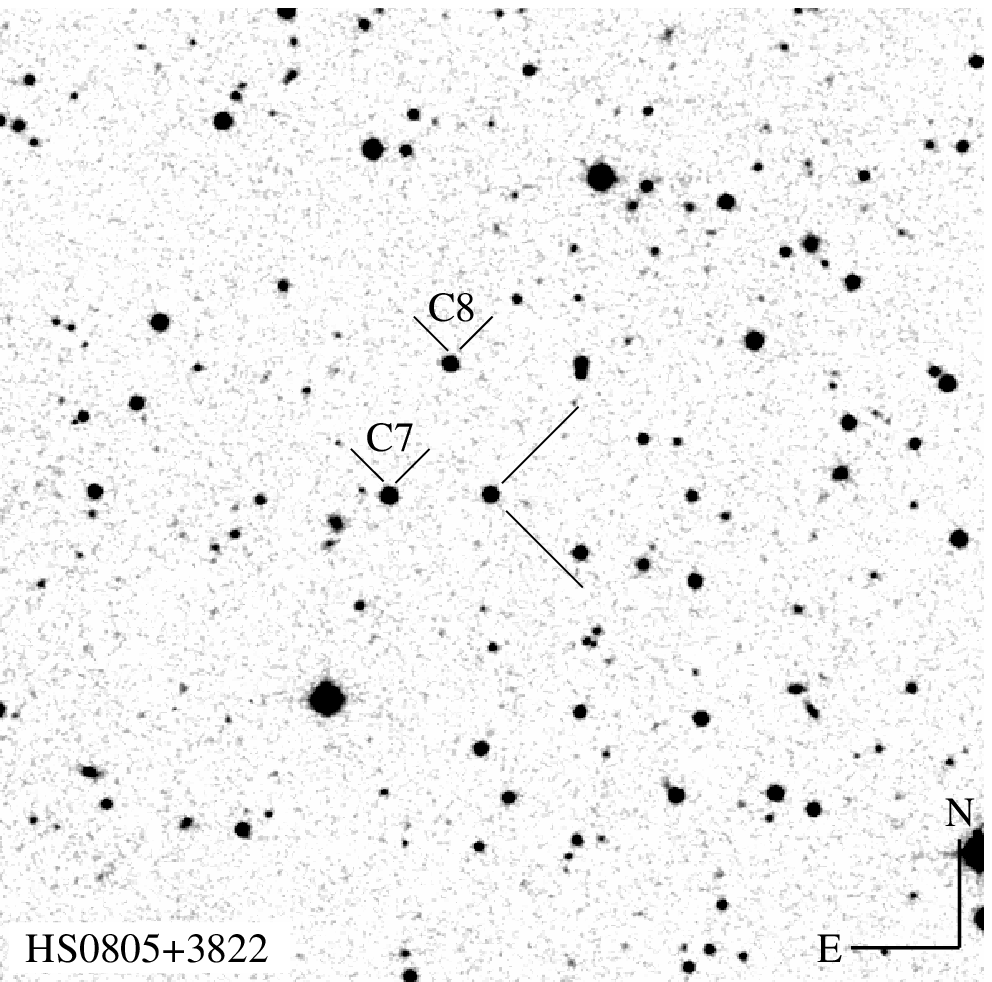}
\hfill
\includegraphics[width=5.8cm]{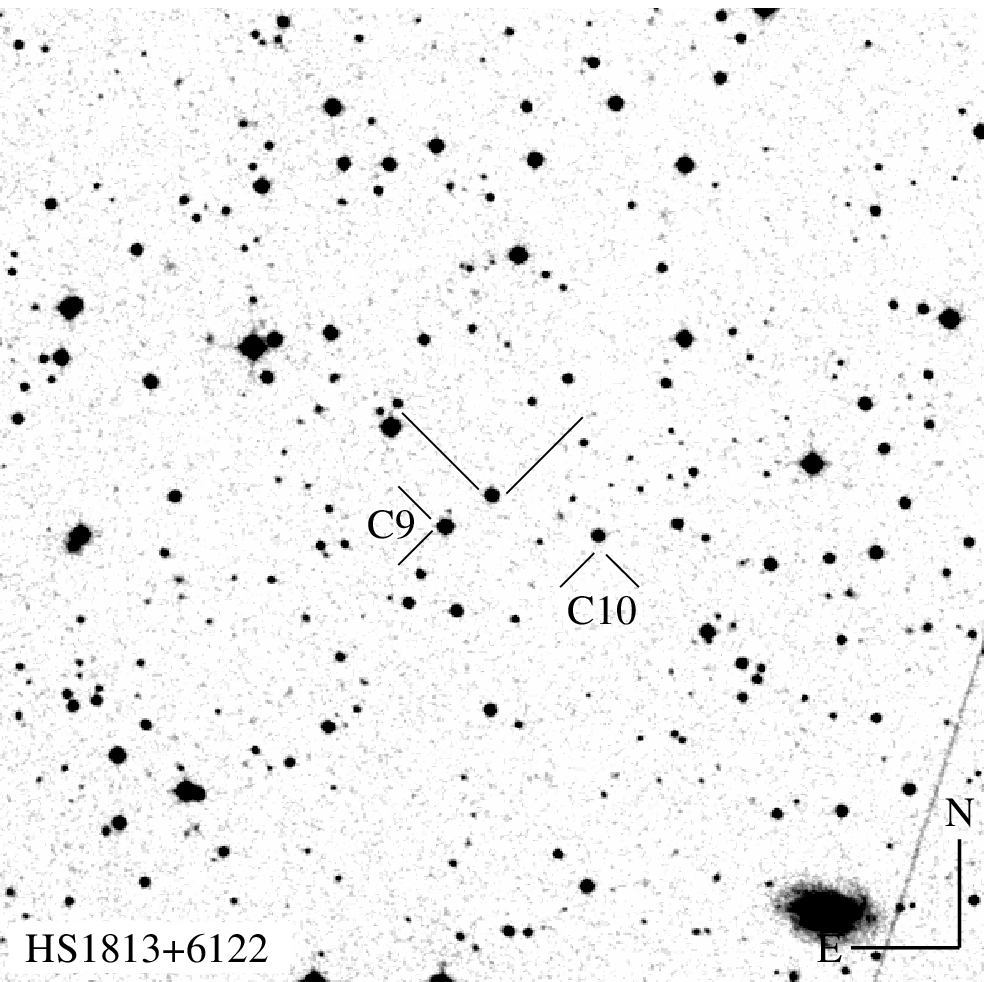}
\caption{\label{fig_fc} $10\arcmin\times10\arcmin$ finding charts for
HS\,0129+2933, HS\,0220+0603, HS\,0455+8315, HS\,0805+3822, and
HS\,1813+6122 obtained from the Digitized Sky Survey.  See
Table\,\ref{t-compstars} for details on the comparison stars C1--C10.}
\end{figure*}

The Palomar-Green (PG) survey (Green, Schmidt \& Liebert \citeyear{greenetal86-1}) led to the
identification of several relatively bright ($V \sim15$), deeply
eclipsing cataclysmic variables (CVs) with orbital periods in the range $3-4$\,h, namely SW\,Sex \citep{penningetal84-1}, DW\,UMa (Shafter, Hessman \& Zhang \citeyear{shafteretal88-1}),
BH\,Lyn \citep{thorstensenetal91-2, dhillonetal92-1}, and PX\,And
\citep{thorstensenetal91-1}. \cite{szkody+piche90-1} and
\cite{thorstensenetal91-1} established a number of common traits
among these systems, including unusually ``V''-shaped eclipse profiles,
the presence of \Line{He}{II}{4686} emission, a substantial orbital phase lag ($\sim 0.2$ cycle) of the radial velocities of the Balmer lines with respect to the motion of the white dwarf, and single-peaked emission lines that display central absorption dips around orbital phases $\simeq0.4-0.7$. The ``SW Sex
phenomenon'' was later extended to lower orbital inclinations after the identification of several grazingly eclipsing and non-eclipsing CVs which exhibit the spectroscopic properties characteristic of the SW Sex class  (WX\,Ari: \citealt{beuermannetal92-1,rodriguez-giletal00-1}; V795\,Her: \citealt{casaresetal96-1}; LS\,Peg: Mart\'\i nez-Pais, Rodr\'\i guez-Gil \& Casares \citeyear{martinez-paisetal99-1}; Taylor, Thorstensen \& Patterson \citeyear{tayloretal99-1}; V442\,Oph: Hoard, Thorstensen \& Szkody \citeyear{hoardetal00-1}).

The observational characteristics of the SW\,Sex stars are not easily
reconciled with the properties of a simple, steady-state hot optically
thick accretion disc which is expected to be found in
intrinsically bright, weakly-magnetic CVs above the period gap. Nevertheless, a variety of mechanisms have been invoked to explain the behaviour of
the SW\,Sex stars, such as stream overflow
\citep{hellier+robinson94-1, hellier96-1}, magnetic white dwarfs
\citep{williams89-1, casaresetal96-1, rodriguez-giletal01-1, hameury+lasota02-1}, magnetic propellers
in the inner disc \citep{horne99-1}, and self-obscuration of the inner
disc \citep{kniggeetal00-1}. While no unambiguous model for their accretion geometry has been found so far, it is becoming increasingly clear that the SW\,Sex
stars are not rare and unusual systems, but represent an important, if not dominant
fraction of all CVs in the orbital period range $3-4$\,h
\citep{rodriguez-gil05-1, aungwerojwitetal05-1}. Hence, a thorough
investigation of this class of systems is important in the context of
understanding CV evolution as a whole.

In this paper, we present five new CVs identified in the Hamburg
Quasar Survey (HQS; \citealt{hagenetal95-1}) which are classified as new SW\,Sex stars on the basis of our follow-up photometry and spectroscopy. The new discoveries bring the total number of confirmed
SW\,Sex stars to 35, and we discuss the global properties of these
objects as a class.

\begin{figure*}
\centerline{\includegraphics[width=10cm]{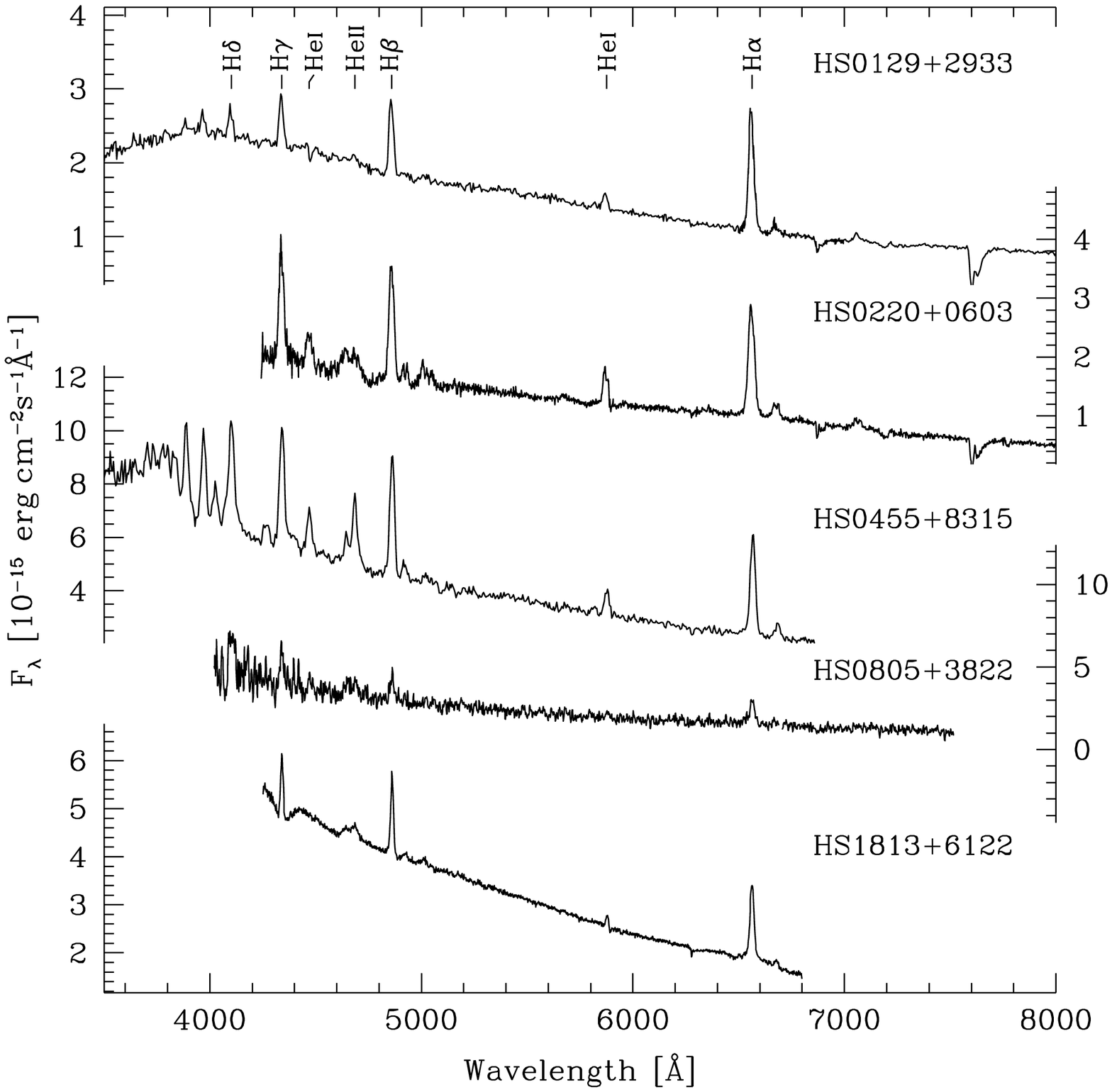}}
\caption{\label{fig_fluxed_spectra} Flux-calibrated optical spectra of the five new CVs.}
\end{figure*}

\begin{table*}
\begin{minipage}{170mm}
\caption{\label{t-targets} Properties of the five new SW\,Sex
  stars. The coordinates and the $B$ and $R$ magnitudes were taken
  from the USNO-B catalog \citep{monetetal03-1mn}. The $J$, $H$, and
  $K_\mathrm{s}$ magnitudes were taken from the 2MASS catalogue. The
  emission line parameters are measured from the public SDSS spectrum.}
\setlength{\tabcolsep}{1.1ex}
\begin{tabular}{lccccc}
\hline\noalign{\smallskip}
    & 
HS\,0129+2933 & 
HS\,0220+0603 & 
HS\,0455+8315 & 
HS\,0805+3822 &
HS\,1813+6122 \\
\hline\noalign{\smallskip}
Right ascension {(J2000)} & 
$01^\mathrm{h}31^\mathrm{m}59.86^\mathrm{s}$ &
$02^\mathrm{h}23^\mathrm{m}01.65^\mathrm{s}$ &
$05^\mathrm{h}06^\mathrm{m}48.27^\mathrm{s}$ &
$08^\mathrm{h}09^\mathrm{m}08.40^\mathrm{s}$ &
$18^\mathrm{h}14^\mathrm{m}29.83^\mathrm{s}$ \\
Declination {(J2000)} &
$+29\degr49\arcmin22.3\arcsec$ &
$+06\degr16\arcmin50.0\arcsec$ &
$+83\degr19\arcmin23.3\arcsec$ &
$+38\degr14\arcmin06.5\arcsec$ &
$+61\degr23\arcmin34.5\arcsec$ \\
Orbital period (h) &
3.35129827(65) &
3.58098501(34) &
3.56937674(26) &
$\simeq 3.22$ &
$\simeq 3.38$  \\
$B/R$ magnitudes  & 
15.3 / 14.7 &
16.5 / 16.2 &
15.2 / 14.6 &
14.7 / 14.6 &
15.1 / 14.5 \\
$J/H/K_\mathrm{s}$ magnitudes  & 
14.6\,/\,14.4\,/\,14.3 &
15.4\,/\,15.1\,/\,14.8 &
14.4\,/\,14.3\,/\,14.1 &
15.3\,/\,15.2\,/\,14.9 &
14.8\,/\,14.7\,/\,14.6 \\
\Ha\ EW [\AA] / FHWM [$\mathrm{km\,s^{-1}}$] &
61 / 1160 &
73 / 1390 &
50 / 1180 &
25 / 1040 &
22 / 1050 \\
\Hb\ EW [\AA] / FHWM [$\mathrm{km\,s^{-1}}$] &
25 / 1265 &
43 / 1560 &
29 / 1250 &
11 /  970 &
 6 / 1020 \\
\Hg\ EW [\AA] / FHWM [$\mathrm{km\,s^{-1}}$] &
16 / 1350 &
25 / 1545 &
31 / 1540 &
10 / 1010 &
5 / 930 \\
\Line{He}{I}{5876} EW [\AA] / FHWM [$\mathrm{km\,s^{-1}}$] &
8 / 1120 &
15 / 1270 &
-- / -- &
2  / 720 &
1 / 720 \\
\Line{He}{I}{6678} EW [\AA] / FHWM [$\mathrm{km\,s^{-1}}$] &
5 / 1090 &
9 / 1240 &
10 / 1240 &
4  / 830 &
-- / -- \\
\Line{He}{II}{4686} EW [\AA] / FHWM [$\mathrm{km\,s^{-1}}$] &
-- / -- &
6 / 2190 &
17 / 1900 &
4  / 700 &
-- / -- \\
\hline\noalign{\smallskip}
\end{tabular}
\end{minipage}
\end{table*}

\section{Observations and data reduction}

\subsection{Identification}
The five new CVs (Fig.\,\ref{fig_fc}) were selected for follow-up
observations upon the detection of emission lines in their HQS
spectra. Identification spectra of HS\,0220+0603, HS\,0805+3822, and
HS\,1813+6122 were respectively obtained in October 1990 with the
Calar Alto 3.5-m telescope, in December 1996 with the 1.5-m
Tillinghast telescope at Fred Lawrence Whipple Observatory, and in
June 1991 with the Calar Alto 2.2-m telescope as part of QSO and
galaxy searches.  HS\,0129+2933 and HS\,0455+8315 were both identified
as CVs in September 2000 using the Calar Alto 2.2-m telescope as part
of a dedicated search for CVs in the HQS \citep{gaensickeetal02-2}.

HS\,0129+2933 (= TT Tri) was already identified as an eclipsing star by
\citet{romano78-1}. At the time of writing this paper we were aware of a multicolour photometric study by Warren, Shafter \& Reed (\citeyear{warrenetal06-1}), in which an eclipse ephemeris is derived for the first time. 

HS\,0805+3822 was independently
found as a CV in the Sloan Digital Sky Survey
(SDSS\,J080908.39+381406.2, \citealt{szkodyetal03-2mn}) and was identified as a SW\,Sex star.

Flux-calibrated, low resolution identification spectra of the five new CVs are shown in Fig.\,\ref{fig_fluxed_spectra}. The spectra are dominated by
strong, single-peaked Balmer and \he{i} emission lines on top of blue continua. The high excitation emission lines of \hel{ii}{4686} and the \Ion{C}{iii}/\Ion{N}{iii} Bowen blend near 4640 \AA~are also observed, and are especially intense in HS 0455+8315. Table~\ref{t-targets} summarises some properties of the new CVs.

\begin{table}
\caption{\label{t-compstars} Comparison stars used for the
  differential CCD photometry (see Fig.\,\ref{fig_fc}).}
\begin{tabular}{cccc}
\hline\noalign{\smallskip}
~~~~ID & ~~~~~~~~~~~~~~USNO-A2.0 &  ~~~~~~$R$ & ~~~~~~$B$~~~ \\
\hline\noalign{\smallskip}
~~~~C01 & ~~~~~~~~~~~~~~1125-00509642 & ~~~~~~13.3 & ~~~~~~14.0~~~ \\
~~~~C02 & ~~~~~~~~~~~~~~1125-00510117 & ~~~~~~14.8 & ~~~~~~15.2~~~ \\
~~~~C03 & ~~~~~~~~~~~~~~0900-00553937 & ~~~~~~15.8 & ~~~~~~17.7~~~ \\
~~~~C04 & ~~~~~~~~~~~~~~0900-00554015 & ~~~~~~16.4 & ~~~~~~18.9~~~ \\
~~~~C05 & ~~~~~~~~~~~~~~1725-00217755 & ~~~~~~13.8 & ~~~~~~15.1~~~ \\
~~~~C06 & ~~~~~~~~~~~~~~1725-00218117 & ~~~~~~15.4 & ~~~~~~15.9~~~ \\
~~~~C07 & ~~~~~~~~~~~~~~1275-07186372 & ~~~~~~15.0 & ~~~~~~15.0~~~ \\
~~~~C08 & ~~~~~~~~~~~~~~1275-07186091 & ~~~~~~15.7 & ~~~~~~17.2~~~ \\
~~~~C09 & ~~~~~~~~~~~~~~1500-06459684 & ~~~~~~14.6 & ~~~~~~15.1~~~ \\
~~~~C10 & ~~~~~~~~~~~~~~1500-06458763 & ~~~~~~15.8 & ~~~~~~16.5~~~ \\
\hline\noalign{\smallskip}
\end{tabular}
\end{table}

\subsection{Optical photometry}

We obtained differential CCD photometry of the five new CVs at seven
different telescopes: the 2.2-m telescope at Calar Alto Observatory, the 1.2-m
telescope at Kryoneri Observatory, the 0.82-m IAC80 and the 1-m
Optical Ground Station at Observatorio del Teide, the 0.7-m telescope
of the Astrophysikalisches Institut Potsdam, the 1.2-m telescope at
Fred Lawrence Whipple Observatory, and the 0.7-m
Schmidt-V\"ais\"al\"a telescope at Tuorla Observatory, and the 0.35-m telescope at West Challow Observatory. Details on the
instrumentation are given in the notes to Table\,\ref{t-obslog}, which also contains the log of the photometric observations. The
data obtained at Calar Alto, Kryoneri, and Tuorla were reduced with
the pipeline described by \citet{gaensickeetal04-1}, which applies
bias and flat-field corrections in \texttt{MIDAS} and then uses
\texttt{Sextractor} \citep{bertin+arnouts96-1} to extract aperture
photometry for all objects in the field of view.  The reduction of the
AIP data was carried out completely using
\texttt{MIDAS}. \texttt{IRAF}\footnote{\texttt{IRAF} is distributued
by the National Optical Astronomy Observatories.} was used to correct the OGS, IAC80, and FLWO data for the bias level and flat-field variations, and to compute Point Spread Function (PSF) magnitudes of the target and comparison stars. Fig.\,\ref{fig_fc} shows finding charts for the five new CVs and
indicates the comparison stars used for differential photometry. The USNO
$R$ and $B$ magnitudes of the comparison stars are given in
Table\,\ref{t-compstars}. Sample light curves are shown in
Fig.\,\ref{fig_sample_lc}. HS\,0129+2933, HS\,0220+0603, and HS\,0455+8315
are deeply eclipsing, HS\,0805+3822 displays evidence for grazing
eclipses (similar to WX\,Ari, \citealt{rodriguez-giletal00-1}), and
all five systems show substantial short time-scale variability. From the scatter in the comparison star light curves we estimate that the differential photometry is accurate to 1 per cent.

\subsection{Optical spectroscopy}

The spectroscopic data were obtained with five different telescopes:
the 2.5-m Isaac Newton Telescope (INT) and the 2.56-m Nordic Optical
Telescope (NOT) on La Palma, the 2.2-m and the 3.5-m telescopes
at Calar Alto Observatory, and the 2.7-m telescope at McDonald Observatory. The log of spectroscopic observations can also be found in
Table~\ref{t-obslog}. Details on the different
telescope/spectrograph setups are given in what follows:

\begin{list}{\arabic{fistro}.}{\usecounter{fistro}}
\item At the INT we used the Intermediate Dispersion Spectrograph
(IDS) with the R632V grating, the $2048 \times 4100$ pixel
EEV10a CCD detector, and a 1.1\arcsec~slit width. With this setup we sampled the wavelength region
$\lambda\lambda4400-6700$ at a resolution (full width at half maximum,
hereafter FWHM) of $\sim 2.5$ \AA.
\item The Andaluc\'\i a Faint Object Spectrograph and Camera (ALFOSC)
was in place at the NOT. The spectra were imaged on the $2048 \times
2048$ pixel EEV chip (CCD \#8). A spectral resolution of $\sim
3.7$ \AA~(FHWM) was achieved by using the grism \#7 (plus the
second-order blocking filter WG345) and a 1\arcsec~slit width. The useful
wavelength interval this configuration provides is
$\lambda\lambda3800-6800$.
\item The spectroscopy at the 2.2-m Calar Alto telescope was
performed with CAFOS. A 1.2\arcsec~slit width and the G--100 grism granted
access to the $\lambda\lambda4200-8300$ range with a resolution of
$\sim 4.5$ \AA~(FWHM) on the standard SITe CCD ($2048 \times 2048$
pixel).
\item The double-armed TWIN spectrograph was used to carry out the
observations at the 3.5-m telescope in Calar Alto. The blue arm was
equipped with the T05 grating, while the T06 grating was in place in
the red arm. The wavelength ranges $\lambda\lambda4070-5160$ and
$\lambda\lambda6080-7180$ were sampled at 1.32 \AA~and 1.23
\AA~resolution (FWHM; 1.5\arcsec~slit width) in the blue and the red,
respectively.
\item The Large Cassegrain Spectrometer (LCS) on the 2.7-m telescope at McDonald Observatory was equipped with grating \#43 and the TI1 $800 \times 800$\,pixel CCD detector. The use of a 1.0\arcsec~slit width resulted in a resolution of 3\,\AA~(FWHM) and a wavelength range of $\lambda\lambda3800-5030$. 
\end{list}

After the effects of bias and flat field structure were removed from
the raw images, the sky background was subtracted. The one-dimensional
target spectra were then obtained using the optimal extraction
algorithm of \cite{horne86-1}. For wavelength calibration, a low-order
polynomial was fitted to the arc data, the \textit{rms} being always
smaller than one tenth of the dispersion in all cases. The
pixel-wavelength correspondence for each target spectrum was obtained
by interpolating between the two nearest arc spectra. These reduction
steps were performed with the standard packages for long-slit spectra
within \texttt{IRAF}.

\subsection{\textit{HST}/STIS far-ultraviolet spectroscopy}
\begin{figure}
\includegraphics[angle=-90,width=\columnwidth]{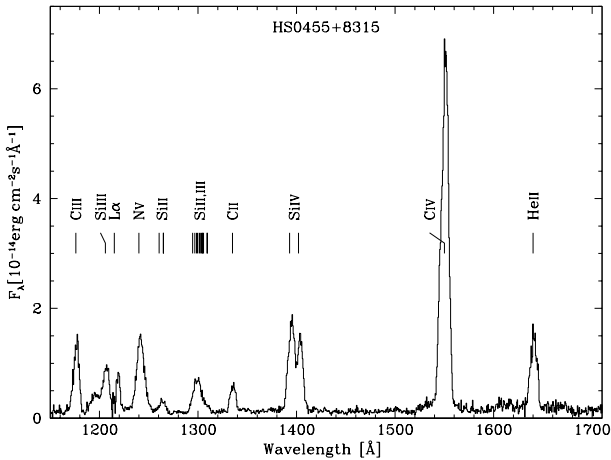}
\caption{\label{fig_hs0455_hst} \textit{HST}/STIS far-ultraviolet
spectrum of HS\,0455+8315.}
\end{figure}

A single far-ultraviolet (FUV) spectrum of HS\,0455+8315 was obtained with
the \textit{Hubble Space Telescope}/Space Telescope Imaging
Spectrograph (\textit{HST}/STIS) as part of a large survey of the FUV
emission of CVs \citep{gaensickeetal03-1}. The data were obtained
using the G140L grating and the $52\arcsec\times0.2\arcsec$ aperture,
resulting in a spectral resolution of $\lambda/\Delta\lambda \simeq 1000$ and a wavelength
coverage of $\lambda\lambda1150-1710$. The spectrum (Fig.~\ref{fig_hs0455_hst}) was obtained at an orbital
phase of $\simeq0.75$, well outside the eclipse (see Sect.~\ref{sect_hst_hs0455} for a discussion). The STIS
acquisition image showed the system at an approximate $R_\mathrm{c}$ magnitude
of $15.4$, close to the normal high-state brightness. 

\begin{table*}
\setlength{\tabcolsep}{0.95ex}
\caption[]{Log of the observations\label{t-obslog}.}
\vspace*{-2.5ex}

\begin{minipage}[t]{8.8cm}
\begin{tabular}[t]{lccccc}
\hline\noalign{\smallskip}
Date & UT &  Telescope & Filter/ & Exp. & Frames  \\    
     &    &            & Grism   & (s)  &          \\    
\hline\noalign{\smallskip}
\multicolumn{6}{l}{\textbf{HS\,0129+2933}} \\
2002 Aug 29 & 03:22-04:35 & INT         &R632V     & 600 & 7\\
2002 Aug 31 & 05:00-05:40 & INT         &R632V     & 600 & 4\\
2002 Sep 02 & 04:20-05:01 & INT         &R632V     & 600 & 4\\
2002 Sep 03 & 04:39-05:20 & INT         &R632V     & 600 & 4\\
2003 Dec 17 & 19:33-01:11 & NOT         &Grism \#7 & 600 & 27 \\
2002 Sep 22 & 23:42-03:35 & KY          &$R$       &  10 & 1035 \\
2003 Sep 29 & 05:29-06:00 & IAC80       & clear    &  15 & 75 \\
2003 Dec 15 & 18:24-23:56 & CA22        & clear    &  15 & 641 \\
2003 Dec 16 & 18:07-19:35 & CA22        & clear    &  20 & 151 \\
2003 Dec 25 & 20:06-21:52 & CA22        & clear    &  15 & 176 \\
2006 Nov 21 & 22:22-23:39 & WCO         & clear    &  60 & 73 \\
2006 Dec 11 & 18:49-19:11 & WCO         & clear    &  60 & 22 \\
2006 Dec 16 & 22:21-23:38 & WCO         & clear    &  60 & 73 \\
\noalign{\smallskip}
\multicolumn {6}{l}{\textbf{HS\,0220+0603}} \\
2002 Oct 15 & 23:13-02:47 & KY          & $R$      &  25 & 328 \\
2002 Oct 17 & 01:13-03:44 & KY          & $R$      &  25 & 279 \\
2002 Dec 08 & 23:27-00:13 & CA22        & G-200    & 600   & 4 \\
2002 Dec 16 & 18:25-19:20 & CA22        & G-200    & 600   & 5 \\
2002 Dec 29 & 20:39-23:41 & CA22        & $V$      &  30 & 217 \\
2002 Dec 31 & 18:30-21:02 & CA22        & $V$      &  30 & 145 \\
2003 Jan 26 & 19:41-23:01 & IAC80       & $V,R$    & 100 & 119 \\
2003 Jan 28 & 19:24-23:24 & IAC80       & $V$      & 120 & 102 \\
2003 Jul 11 & 04:19-05:10 & OGS         & clear    &  50 & 54 \\
2003 Jul 13 & 04:08-05:25 & OGS         & clear    &  30 & 119 \\
2003 Sep 22 & 03:07-04:16 & IAC80       & clear    &  30 & 100 \\
2003 Sep 29 & 00:07-01:24 & IAC80       & clear    &  30 & 113 \\
2003 Nov 09 & 00:50-02:03 & IAC80       & $R$      &  60 &  58 \\
2003 Nov 17 & 23:41-00:27 & OGS         & clear    &  17 & 108 \\
2003 Dec 15 & 19:28-21:04 & NOT         & Grism \#7 & 600 & 9 \\
2003 Dec 16 & 19:26-02:33 & NOT         & Grism \#7 & 600 & 38 \\
2006 Nov 21 & 19:15-20:11 & WCO         & clear    &  60 & 55 \\
2006 Dec 11 & 19:16-19:49 & WCO         & clear    &  60 & 33 \\
2006 Dec 11 & 22:47-23:23 & WCO         & clear    &  60 & 36 \\
2006 Dec 16 & 21:04-22:16 & WCO         & clear    &  60 & 70 \\
\noalign{\smallskip}
\multicolumn{6}{l}{\textbf{HS\,0455+8315}}\\
2000 Oct 20 & 18:57-21:44 & AIP         & $R$   &  60   & 142 \\
2000 Nov 10 & 17:30-21:48 & AIP         & $R$   &  30   & 364 \\
2000 Nov 16 & 16:45-21:14 & AIP         & $R$   &  30   & 425 \\
2000 Dec 04 & 17:09-17:50 & AIP         & $R,B$ &  30   &  64 \\
2000 Dec 05 & 16:39-18:03 & AIP         & $R,B$ &  30   & 135 \\
2000 Jan 01 & 03:30-05:31 & CA35        & T05/T06 & 600 & 11 \\
2001 Jan 01 & 23:03-23:45 & CA35        & T05/T06 & 300 & 5 \\
2001 Jan 02 & 00:32-04:38 & CA35        & T05/T06 & 400 & 33 \\
2002 Dec 09 & 09:06       & HST         & G140L & 730   & 1 \\
2006 Nov 21 & 20:30-22:02 & WCO         & clear &  60 & 88 \\
2006 Nov 23 & 23:05-00:09 & WCO         & clear &  60 & 62 \\
2006 Dec 08 & 20:05-21:03 & WCO         & clear &  50 & 68 \\
\noalign{\smallskip}\hline
\end{tabular}
\end{minipage}
\hfill
\begin{minipage}[t]{8.8cm}
\begin{tabular}[t]{lccccc}
\hline\noalign{\smallskip}
Date & UT &  Telescope & Filter/ & Exp. & Frames  \\    
     &    &            & Grism   & (s)  &          \\    
\hline\noalign{\smallskip}
2007 Jan 11 & 21:28-22:40 & WCO         & clear &  60 & 70 \\
2007 Jan 13 & 23:33-00:15 & WCO         & clear &  50 & 50 \\
2007 Jan 14 & 21:01-21:52 & WCO         & clear &  50 & 60 \\[1ex]
\noalign{\smallskip}
\multicolumn{6}{l}{\textbf{HS0805+3822}} \\
2005 Feb 11 & 19:28-04:32 & Tuorla  & clear &  90    & 325 \\
2005 Feb 15 & 18:36-01:03 & Tuorla  & clear &  80    & 257 \\
2005 Feb 28 & 04:39-10:00 & FLWO1.2 & clear &  15-20 & 769 \\
2005 Mar 01 & 03:12-09:56 & FLWO1.2 & clear &  15    & 784 \\
2005 Mar 04 & 05:59-09:33 & FLWO1.2 & clear &  15-25 & 402 \\
2005 Mar 24 & 20:03-00:04 & Tuorla  & clear &  30    & 373 \\
2005 Mar 25 & 20:00-02:21 & Tuorla  & clear &  30    & 608 \\
2005 Mar 27 & 03:07-08:18 & FLWO1.2 & clear &  15-20 & 674 \\
2005 Mar 28 & 04:06-07:12 & FLWO1.2 & clear &  15-20 & 390 \\
2005 Mar 29 & 04:48-08:04 & FLWO1.2 & clear &  15-30 & 321 \\[1ex]
\noalign{\smallskip}
\multicolumn{6}{l}{\textbf{HS1813+6122}}\\
2000 Sep 24 & 20:23-20:54 & CA22    & R-100  & 600   &   2 \\
2001 Aug 23 & 21:46-01:26 & AIP     & $R$    &  30   & 346 \\
2001 Aug 23 & 20:10-01:05 & AIP     & $R$    &  60   & 251 \\
2001 Aug 23 & 19:58-00:20 & AIP     & $R$    &  60   & 199 \\
2002 Jul 03 & 23:47-02:29 & KY      & $R$    & 120   &  67 \\
2002 Jul 05 & 19:54-22:31 & KY      & $R$    & 120   &  69 \\
2002 Aug 22 & 18:44-21:36 & KY      & clear  &  10   & 660 \\
2002 Aug 23 & 18:59-22:21 & KY      & clear  &  10   & 800 \\
2002 Sep 04 & 21:24-23:58 & KY      & $R$    &  10   & 516 \\
2002 Sep 04 & 00:23-00:54 & INT     & R632V  & 600   &   4 \\
2002 Sep 06 & 18:34-21:57 & KY      & $R$    &  35   & 276 \\
2003 Aug 20 & 18:11-21:43 & KY      & clear  &  10   & 649 \\
2002 Aug 27 & 22:38-00:11 & INT     & R632V  & 600   &   9 \\
2002 Aug 31 & 23:37-00:28 & INT     & R632V  & 600   &   5 \\
2002 Sep 01 & 21:55-22:37 & INT     & R632V  & 600   &   4 \\
2002 Sep 03 & 20:26-21:07 & INT     & R632V  & 600   &   4 \\
2003 Jun 28 & 22:49-00:35 & KY      & $R$    & 60    &  93 \\
2003 Jul 15 & 02:29-05:09 & OGS     & clear  & 15    & 442 \\
2003 Sep 23 & 20:23-00:05 & IAC80   & clear  &  7    & 690 \\
2003 Sep 24 & 20:02-22:50 & IAC80   & clear  &  7    & 502 \\
2003 Sep 25 & 21:18-23:48 & IAC80   & clear  & 10    & 400 \\
2003 Sep 26 & 19:53-21:08 & IAC80   & clear  &  7    & 248 \\
2003 Sep 27 & 19:36-23:19 & IAC80   & clear  &  7    & 680 \\
2003 May 18 & 01:00-04:08 & INT     & R632V  & 600   &   6 \\
2003 May 19 & 02:48-04:13 & INT     & R632V  & 600   &   8 \\
2004 May 24 & 02:13-03:36 & IAC80   & clear  &  15   & 134 \\
2004 May 25 & 01:44-05:19 & IAC80   & clear  &  15   & 411 \\
2004 May 26 & 00:34-05:15 & IAC80   & clear  &  15   & 720 \\
2003 Jun 29 & 00:52-03:58 & CA22    & G--100 & 600   &  15 \\
2003 Jun 30 & 03:13-04:11 & CA22    & G--100 & 600   &   5 \\
2004 Jul 17 & 04:17-04:50 & McD     & \#43 & 600   &   3 \\
2004 Jul 18 & 04:11-05:19 & McD     & \#43 & 600   &   6 \\[1ex]
\noalign{\smallskip}\hline
\end{tabular}
\end{minipage}

\smallskip
\begin{minipage}{168mm}
Notes on the instrumentation used for CCD photometry. CA22: 2.2-m
telescope at Calar Alto Observatory, using CAFOS with a
$2\mathrm{k}\times2\mathrm{k}$ pixel SITe CCD; KY: 1.2-m telescope at
Kryoneri Observatory, using a Photometrics SI-502 $516\times516$ pixel
CCD camera; IAC80: 0.82-m telescope at Observatorio del Teide,
equipped with Thomson $1\mathrm{k}\times1\mathrm{k}$ pixel CCD camera;
OGS: 1-m Optical Ground Station at Observatorio del Teide, equipped
with Thomson $1\mathrm{k}\times1\mathrm{k}$ pixel CCD camera; AIP:
0.7-m telescope of the Astrophysikalisches Institut Potsdam, with
$1\mathrm{k}\times1\mathrm{k}$ pixel SITe CCD; FLWO: 1.2-m telescope
at Fred Lawrence Whipple Observatory, equipped with MINICAM containing
three $2048\times4608$ EEV CCDs; Tuorla: 0.7-m Schmidt-V\"ais\"al\"a
telescope at Tuorla Observatory, equipped with a SBIG ST--8 CCD
camera; WCO: 0.35-m telescope at West Challow Observatory with SXV-H9 CCD camera.
\end{minipage}
\end{table*}

\begin{table}
\caption[]{\label{t-mideclipse}Eclipse timings (given in
  $\mathrm{HJD}-2450000$), cycle number, and the difference in
  observed minus computed eclipse times using the ephemerides in
  Eqs.~(\ref{e-hs0129}--\ref{e-hs0805_1}). See \citet{warrenetal06-1} for additional eclipse timings of HS\,0129+2933.}

\setlength{\tabcolsep}{0.95ex}
\begin{flushleft}
\begin{tabular}{rrrrrr}
\hline\noalign{\smallskip}
\multicolumn{1}{c}{$T_0$} & \multicolumn{1}{c}{Cycle} &  \multicolumn{1}{c}{$O-C$~(s)} &   
\multicolumn{1}{c}{$T_0$} & \multicolumn{1}{c}{Cycle} &  \multicolumn{1}{c}{$O-C$~(s)} \\   
\hline\noalign{\smallskip}
\noalign{\smallskip}
\multicolumn{3}{l}{\textbf{HS\,0129+2933}} & 2961.51100 & 2667  &  --1  \\
2540.53210 &  0    & --30   		   & 4061.32113 & 10038 &   10  \\  
2989.32739 &  3214 &   22   		   & 4081.31481 & 10172 &  --4	\\	  
2989.46702 &  3215 &   21   		   & 4081.46397 & 10173 &  --8	\\	  
2990.30418 &  3221 & --36   		   & 4086.38789 & 10206 &  --2	\\	
2999.38161 &  3286 &   50   		   & \multicolumn{3}{l}{\textbf{HS\,0455+8315}}	\\
4061.46337 & 10892 &    5                  & 1859.24683 &     0 &  --7  \\	       	
4081.29206 & 11034 &   20                  & 1859.39549 &     1 & --13  \\		       	
4086.45840 & 11071 &  --1                  & 1865.34449 &    41 & --12  \\		     	
\multicolumn{3}{l}{\textbf{HS\,0220+0603}} & 1884.23301 &   168 &    8  \\		     	
2563.57452 &   0  &    42 		   & 1884.23333 &   168 &   24  \\		     	  
2564.61852 &   7  &     2 		   & 4061.40138 & 14807 & --28  \\		  
2638.47610 &  502 &  --18 		   & 4063.48339 & 14821 & --13  \\		  
2666.37811 &  689 &   --3  		   & 4078.35639 & 14921 &   43  \\		
2666.37818 &  689 &     3  		   & 4112.41340 & 15150 & --25  \\		
2668.31751 &  702 &  --29 		   & 4114.49583 & 15164 &    0  \\		  
2668.46682 &  703 &  --20 		   & 4115.38843 & 15170 &   22  \\		
2831.70004 & 1797 &  --21 		   & \multicolumn{3}{l}{\textbf{HS\,0805+3822}} \\     
2834.68439 & 1817 &   --5  		   & 3413.42628 &  0   &   --9 \\	     	
2904.66298 & 2286 &    10                  & 3455.38312 &  313 &   231 \\		    
2911.52646 & 2332 &     4                  & 3455.51361 &  314 &  --75 \\		     	
2952.55900 & 2607 &    40                  & 3457.79144 &  331 & --147 \\                     
\noalign{\smallskip}\hline
\end{tabular}
\end{flushleft}
\end{table}

\section{Photometric periods}

\subsection{HS\,0129+2933, HS\,0220+0603, and HS\,0455+8315} 
The deep eclipses detected in the light curves of HS\,0129+2933,
HS\,0220+0603, and HS\,0455+8315 provide accurate information on the
orbital periods of the systems. Determining the time of mid-eclipse\footnote{The time of mid-eclipse, $T_0$, is the time of inferior conjunction of the donor star.} in
SW\,Sex stars is notoriously difficult due to the asymmetric shape of
the eclipse profiles.  We have therefore employed the following method: the
observed eclipse profile is mirrored in time around an estimate of the
eclipse centre, and the mirrored profile is overplotted on the
original eclipse data. The time of mid-eclipse is then varied until
the central part of both eclipse profiles overlap as closely as
possible. This empirical method proved to be somewhat more robust
compared to e.g. fitting a parabola to the eclipse profile. The
mid-eclipse times are reported in Table\,\ref{t-mideclipse}. An
initial estimate of the cycle count was then obtained by fitting
eclipse phases
$(\phi_0^{\mathrm{observed}}-\phi_0^\mathrm{\mathrm{fit}})^{-2}$ over
a wide range of trial periods (Fig.\,\ref{fig_tsa}). Once an unambiguous
cycle count was established, a linear eclipse ephemeris was fitted to
the times of mid-eclipse. For HS\,0129+2933, we also included the 22 eclipse timings reported by \cite{warrenetal06-1} in our calculations. The resulting ephemerides are:

\begin{equation}
T_0(\mathrm{HJD})=2\,452\,540.53244(21)+0.139637428(27)\times E
\label{e-hs0129}
\end{equation}

for HS\,0129+2933, i.e. $\Porb=3.35129827(65))$\,h, 

\begin{equation}
T_0(\mathrm{HJD})=2\,452\,563.574036(73)+0.149207709(14)\times E
\label{e-hs0220}
\end{equation}

for HS\,0220+0603,  i.e. $\Porb=3.58098501(34)$\,h, and

\begin{equation}
T_0(\mathrm{HJD})=2\,451\,859.24463(12)+0.148724030(11)\times E
\label{e-hs0455}
\end{equation}

for HS\,0455+8315,  i.e. $\Porb=3.56937674(26)$\,h.

\subsection{HS\,0805+3822}
\begin{figure}
\includegraphics[width=8.9cm]{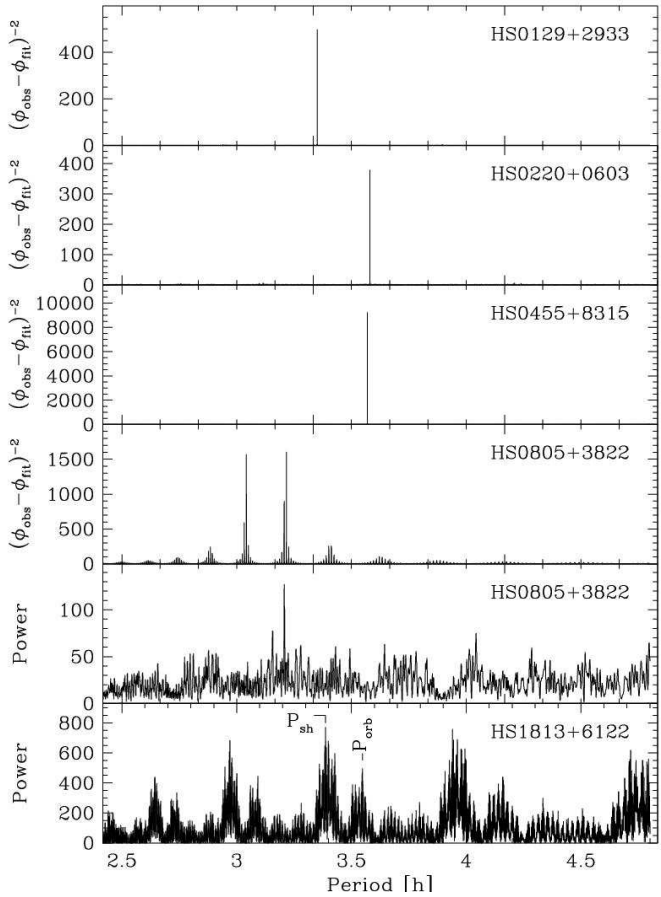}
\caption{\label{fig_tsa} Time series analysis of the CCD photometry of
  the five new SW\,Sex stars. The four top panels show the results of
  eclipse-phase fits for the deeply eclipsing systems HS\,0129+2933,
  HS\,0220+0603, and HS\,0455+8315, as well as for the grazing
  eclipser HS\,0805+3822. For HS\,0805+3822, an AOV periodogram
  \citep{schwarzenberg-czerny96-1} is shown in the second-lowest
  panel, and a \citet{scargle82-1} periodogram is shown for
  HS\,1813+6122 in the bottom panel, with the likely orbital
  period and superhump period indicated.}
\end{figure}

\begin{figure*}
\includegraphics[angle=-90,width=18cm]{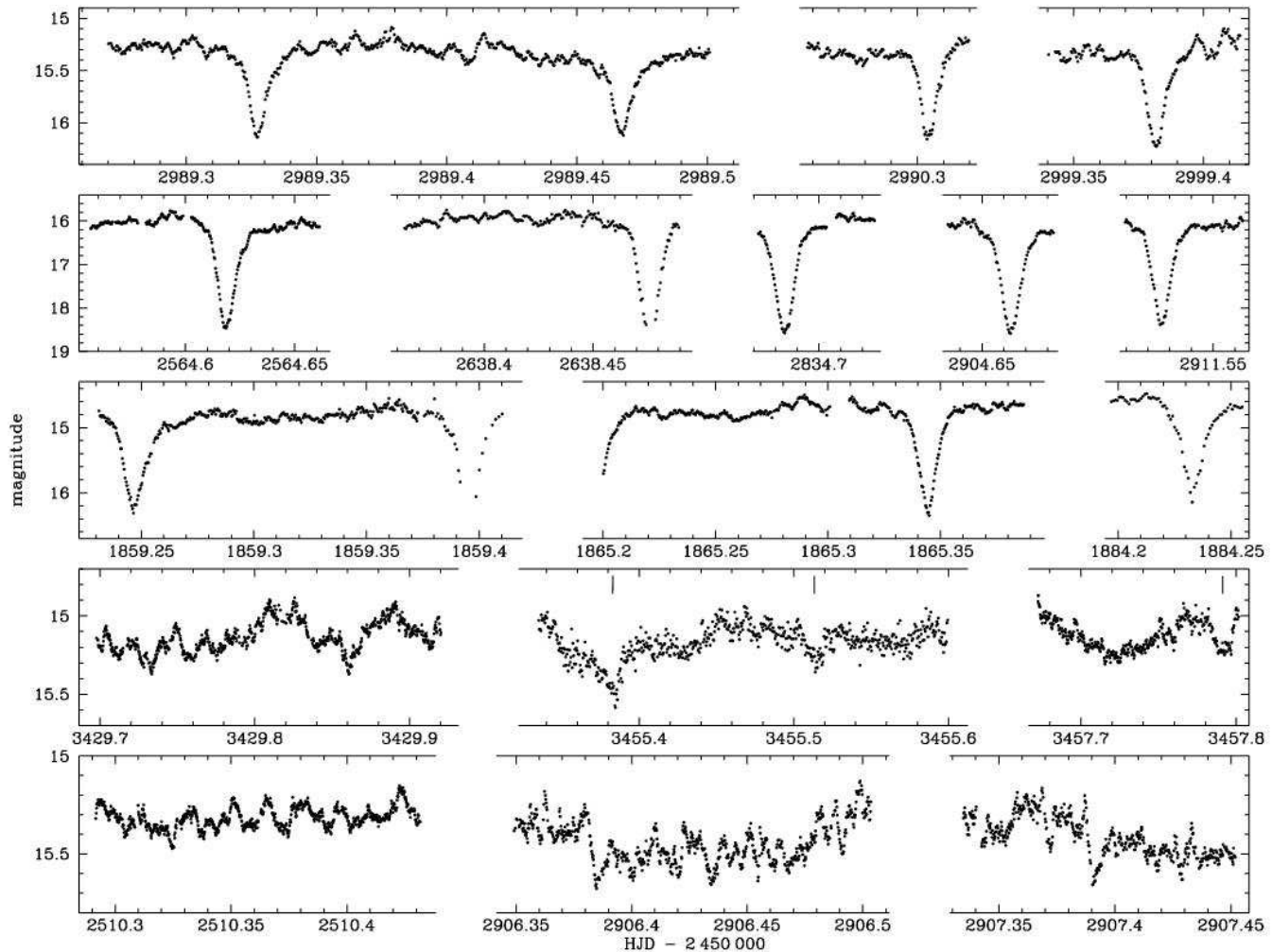}
\caption{\label{fig_sample_lc} Sample light curves of the five new
  SW\,Sex stars. From \textit{top} to \textit{bottom}: HS\,0129+2933, HS\,0220+0603,
  HS\,0455+8315, HS\,0805+3822, and HS\,1813+6122. HS\,0805+3822
  displays grazing eclipses. Those indicated by ticks above the light
  curves were used to calculate an ephemeris.}
\end{figure*}

Some of the light curves of HS\,0805+3822 contain broad dips that we interpret as grazing eclipses, similar to those detected
in WX\,Ari \citep{rodriguez-giletal00-1}. The system displays a
varying level of short time-scale variability, and we restrict the
identification of the (presumed) eclipses to the nights with low
flickering activity (Fig.\,\ref{fig_sample_lc},
Table\,\ref{t-mideclipse}). No unique cycle count can be determined
from the four eclipses detected (Fig.\,\ref{fig_tsa}), and the eclipse
ephemerides determined for the two most likely cycle count aliases are

\begin{equation}
T_0(\mathrm{HJD})=2453413.4264(22)+0.1267611(75)\times E~,
\label{e-hs0805_1}
\end{equation}

i.e. $\Porb=3.0423(2)$\,h, and

\begin{equation}
T_0(\mathrm{HJD})=2453413.4264(23)+0.1340385(84)\times E~,
\label{e-hs0805_2}
\end{equation}

i.e. $\Porb=3.2169(2)$\,h.

\begin{figure}
\includegraphics[width=8.9cm]{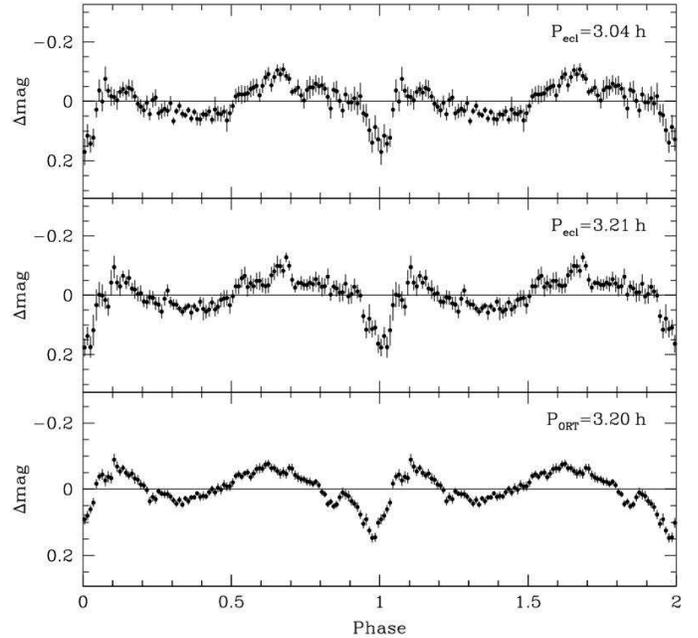}
\caption{\label{fig_hs0805_folded} Phase-folded light curves of
  HS\,0805+3822. \textit{Top} and \textit{middle} panels: data from the nights with low
  flickering activity folded over the eclipse ephemerides given by
  Eqs.\,(\ref{e-hs0805_1},\ref{e-hs0805_2}). \textit{Bottom panel}: all data folded over
  the period determined from an analysis-of-variance periodogram (ORT, \citealt{schwarzenberg-czerny96-1}).}
\end{figure}

~\\
The data of the nights during which
grazing eclipses were detected folded over both periods are shown in
Fig.\,\ref{fig_hs0805_folded}. 

As an alternative approach, we have subjected the combined data from
all nights to an analysis-of-variance (after subtracting the nightly
mean magnitudes) using \citeauthor{schwarzenberg-czerny96-1}'s (1996)
method, and find the strongest signal at 3.21\,h
(Fig.\,\ref{fig_tsa}). Folding all data on that period gives a light
curve which broadly resembles the ``eclipse'' light curves mentioned
above. We conclude that the orbital period of HS\,0805+3822 is
$\simeq 3.22$\,h.

\subsection{HS\,1813+6122\label{sec-hs1813-phot}}
\begin{figure}
\includegraphics[width=8.9cm]{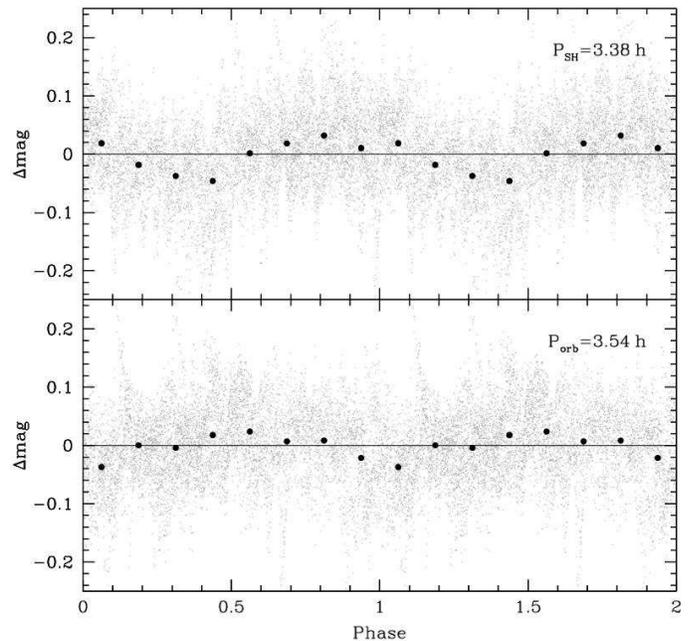}
\caption{\label{fig_hs1813_folded} Phase-folded light curves of
  HS\,1813+6122. \textit{Top panel}: all data folded over
  $P_\mathrm{SH}=3.38$\,h, the strongest signal detected in the
  periodogram (Fig.\,\ref{fig_tsa}), after subtracting a sine fit with
  $P_\mathrm{orb}=3.54$\,h, the period of the neighbouring signal. 
  \textit{Bottom panel}: all data folded over $P_\mathrm{orb}=3.54$\,h after
  subtracting a sine fit with $P_\mathrm{SH}=3.38$\,h. We interpret
  $P_\mathrm{SH}$ and  $P_\mathrm{orb}$ as the superhump and orbital
  periods, respectively.}
\end{figure}

The light curve of HS\,1813+6122 is characterized by rapid
($10-20$\,min) oscillations with a typical amplitude of $0.1-0.2$\,mag,
superimposed by modulations with time-scales of hours and amplitudes
of $\simeq0.1$\,mag. We combined all data after subtracting the
nightly means, and calculated a \cite{scargle82-1} periodogram
(Fig.\,\ref{fig_tsa}, bottom panel). Two broad clusters of signals are
found at 3.38 and 3.54\,h, with the stronger signal at the
shorter period. The same double-cluster structure repeats with lower
amplitudes at several $\pm1$ cycle/day
aliases. Fig.\,\ref{fig_hs1813_folded} shows the photometric data
folded on either period, after subtracting a sine fit with the respective
other period.

Many of the known SW\,Sex stars display positive and/or negative
superhumps in their light curves (e.g. \citealt{patterson+skillman94-1}; Rolfe, Haswell \& Patterson \citeyear{rolfeetal00-1}; \citealt{stanishevetal02-1}; \citealt{pattersonetal02-1}; \citealt{pattersonetal05-3}), and the pattern observed in HS\,1813+6122 fits
into that picture. On the basis of photometry alone, it is usually not
possible to unambiguously identify orbital and superhump periods, but
based on the radial velocity study described below in
Sect.\,\ref{s-hs1813_spectra}, we suggest that $P_\mathrm{SH}=3.38$\,h and
$\Porb=3.54$\,h. 

\section{Spectroscopic analysis}




\subsection{An overabundance of nitrogen in HS\,0220+0603}

\begin{figure}
\centering \includegraphics[width=9cm]{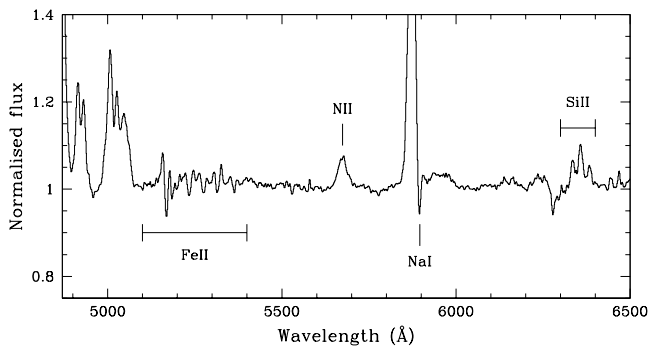}
\caption{\label{fig_hs0220_oddlines} Orbitally-averaged spectrum of
  HS\,0220+0603. Notice the unusual \Line{N}{ii}{5680} emission
  line.}
\end{figure}

HS\,0220+0603 shows emission lines not usually seen in CVs. In
Fig.~\ref{fig_hs0220_oddlines} we plot the average spectrum covering
the region $\lambda\lambda4870-6500$. We identify a group of lines
around $\lambda$5250 as \Ion{Fe}{ii} transitions. They very
likely are emission profiles plus an absorption component since a deep
absorption trough is observed at the position of
\Line{Fe}{ii}{5169}. The \Line{He}{i}{5016} line is abnormally broad
and has a strange profile with three peaks. This is probably the
effect of blended \Ion{He}{i} and \Ion{Fe}{ii} emission (at
$\lambda5018$). But the most notorious feature is the broad
\Line{N}{ii}{5680} emission.  Another triple-peaked emission line
profile is found at $\lambda$6350 which we identify as
\Ion{Si}{ii} emission. The \Line{N}{ii}{5680} transition has been
observed in the intermediate polar HS\,0943+1404
\citep{rodriguez-giletal05-2}, and we interpret its presence as
evidence of an anomalously large abundance of nitrogen.

An observed overabundance of nitrogen in the accretion flow is directly linked to the chemical abundances of the donor star, and suggests that the donor
has undergone nuclear evolution via the CNO cycle. This implies that
the donor had an initial mass of $\ga 1.2\,\mathrm{M}_\odot$ and the
system evolved through a phase of thermal-time scale mass transfer
(\citealt{schenkeretal02-1}; Podsiadlowski, Han \& Rappaport \citeyear{podsiadlowskietal03-1}). The theoretical
models predict that up to 1/3 of all CVs may have undergone nuclear
evolution. This seems to be confirmed by the substantial number of systems with evolved donor stars that have been found (e.g. Jameson, King \& Sherrington \citeyear{jamesonetal80-1};
\citealt{bonnet-bidaud+mouchet87-1}; Thorstensen et al. 2002a, 2002b; \citealt{gaensickeetal03-1})\nocite{thorstensenetal02-1,thorstensenetal02-2}.

\subsection{Radial velocities}

\begin{figure}
\centering
\includegraphics[width=\columnwidth]{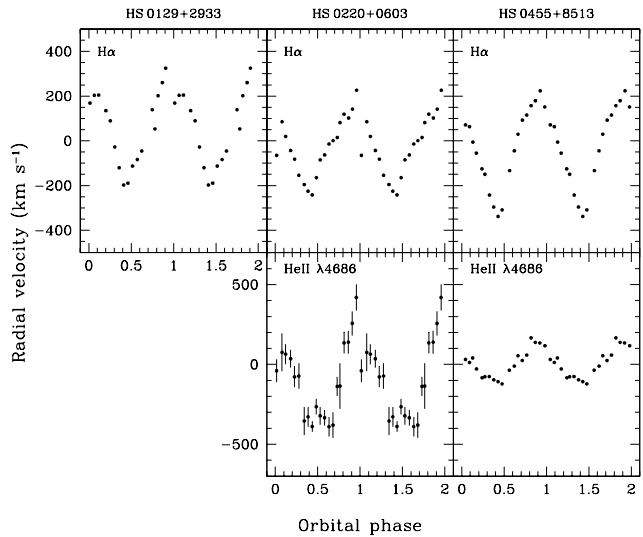}
\caption{\label{fig_vel} Radial velocity curves of the deeply-eclipsing systems HS\,0129+2933, HS\,0220+0603, and HS\,0455+8315 folded on their respective orbital
periods after averaging the data into 20 phase bins. The \Ha~velocities are delayed by $\sim 0.1-0.2$ orbital cycle with respect to the photometric
ephemerides. The same is observed for the \hel{ii}{4686} lines in HS\,0220+0603 and HS\,0455+8315. The orbital cycle has been plotted twice.}
\end{figure}

Radial velocity curves of the \Ha~and \hel{ii}{4686} emission lines were computed for the eclipsing systems, except for HS\,0129+2933 where \hel{ii}{4686} is too weak and only \Ha~was measured. The individual velocity points were obtained by cross-correlating the individual profiles with single Gaussian templates matching the FWHMs of the respective average line profiles (see Table~\ref{t-targets}). Before measuring the velocities, the
normalised spectra were re-binned to a uniform velocity scale centred
on the rest wavelength of each line. The radial velocity curves of the three
deeply-eclipsing systems folded on their respective orbital periods are
presented in Fig.~\ref{fig_vel}.


We fitted sinusoidal functions of the form:$$V_\mathrm{r}=\gamma-K \sin \left[
2\pi \left( \varphi-\varphi_0 \right) \right]$$

\noindent
to the radial velocity curves. The fitting parameters are shown in
Table~\ref{tab_rvcfit}. Note that the $\gamma$ and $K$ parameters given in the table are not the actual systemic velocity and radial velocity amplitude of the white dwarf ($K_1$). The \Ha~line in HS\,0129+2933, HS\,0220+0603, and HS\,0455+8315 is delayed with respect to the motion of the white dwarf by $\varphi_0 \sim 0.2$. The \hel{ii}{4686} radial velocity curves of HS\,0220+0603 and HS\,0455+8315 show the same phase offset. This phase lag indicates that the main emission site is at an angle ($\sim 72\degr$) to the
line of centres between the centre of mass of the binary and the white dwarf. The radial velocity curves are not sinusoidal in shape, showing
significant distortion mainly at phases 0 (mid-eclipse) and 0.5. The
spikes at $\varphi \sim 0$ are due to a rotational disturbance, caused
by the fact that the secondary first occults the disc material
approaching us and then the receding material. This translates into a
red velocity spike before mid-eclipse and a blue spike after it.

\begin{table}
\caption[]{\label{tab_rvcfit} Radial velocity curves fitting parameters.}
\begin{flushleft}
\begin{tabular}{lccc}
\hline\noalign{\smallskip}
Line~~~~~~~~~~~~ & $\gamma$ (\kms) & $K$ (\kms) & $\varphi_0$   \\
\hline\noalign{\smallskip}
\multicolumn{4}{l}{\bf HS\,0129+2933} \\

\Ha  & $69 \pm 1$ & $282 \pm 1$ & $0.240 \pm 0.001$  \\
%
\noalign{\smallskip}\hline
\noalign{\smallskip}
\multicolumn{4}{l}{\bf HS\,0220+0603} \\
\Ha  & $-37.0 \pm 0.7$ & $186 \pm 1$ & $0.170 \pm 0.001$  \\
\Line{He}{II}{4686} & $-88 \pm 16$ & $295 \pm 22$ & $0.24 \pm 0.01$ \\
\noalign{\smallskip}\hline
\noalign{\smallskip}

\multicolumn{4}{l}{\bf HS\,0455+8513} \\

\Ha  & $-27 \pm 1$ & $238 \pm 1$ & $0.155 \pm 0.001$  \\
\Line{He}{II}{4686} & $5 \pm 1$ & $125 \pm 7$ & $0.14 \pm 0.01$ \\
%

%
\noalign{\smallskip}\hline
\end{tabular}
\end{flushleft}
\rmfamily
\end{table}

\subsection{\label{s-hs1813_spectra}The orbital period of HS\,1813+6122}

\begin{figure}
\centering \includegraphics[width=9cm]{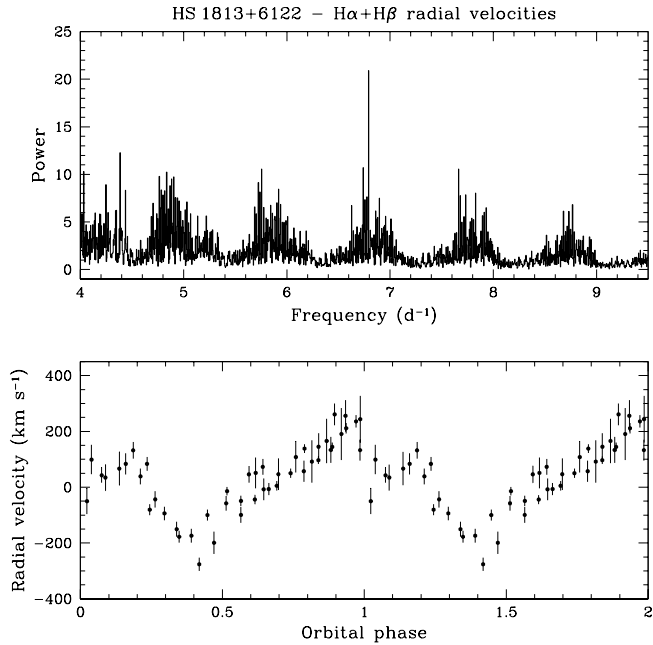}
\caption{\label{fig_havelscargle} \textit{Top}: Scargle periodogram of the \Ha+\Hb~radial velocity curves of HS\,1813+6122. \textit{Bottom}: phase folded radial velocity curve (no phase binning applied). See text for details on the adopted $T_0$. A full cycle has been repeated.}
\end{figure}

Of the five new CVs, HS\,1813+6122 is the only one not showing eclipses, but the photometry suggests a possible orbital period of 3.54 h (Sect.~\ref{sec-hs1813-phot}). In an attempt to confirm this value we performed a period analysis on the \Ha~and \Hb~radial velocities. The velocities were
derived by using the double Gaussian method of \cite{schneider+young80-2} (which gave better results than the single Gaussian technique), adopting a Gaussian FWHM of 200 \kms~and a separation of 1600\,\kms. The 2002 September 1 and 2003 May INT data were obtained under poor observing conditions and were therefore excluded from the analysis. In Fig.~\ref{fig_havelscargle} we show the resulting analysis-of-variance periodogram computed from the combined \Ha~and \Hb~radial velocity curves. The periodogram shows its strongest peak at a period of 3.53\,h,
which confirms the value given by the photometric data as well as the
presence of a negative superhump in the light curve of
HS\,1813+6122. A sine fit to the longest spectroscopic run (2003 Jun
29) with the period fixed at the above value yielded a tentative
zero-phase time of $T_0(\mathrm{HJD})=2\,452\,819.597(2)$. A
preliminary trailed spectrum revealed the characteristic
SW\,Sex high-velocity S-wave reaching its maximum blue velocity at
relative phase $\sim 0.3$ (the phases are defined by using the above $T_0$). By analogy with the eclipsing systems presented in this paper
(Sect.~\ref{sec_trailed}), and with the eclipsing SW Sex stars in general \citep[see e.g.][]{hellier96-1,hoard+szkody97-1,hellier00-1,rodriguez-giletal01-2}, this is expected to happen at absolute phase $\varphi \sim 0.5$. Therefore, the Balmer radial velocities of HS\,1813+6122 are delayed by $\varphi_0 \sim 0.2$ with respect to the white dwarf motion, a defining characteristic of the SW\,Sex stars. Correcting for this we get a new time of zero phase of
$T_0(\mathrm{HJD})=2\,452\,819.568(2)$. Fig.~\ref{fig_havelscargle}
shows the combined \Ha~and \Hb~velocities folded on the orbital
period with the new phase definition.

\subsection{Trailed spectra \label{sec_trailed}}

\begin{figure*}
\centering
  \includegraphics[width=6cm,angle=-90]{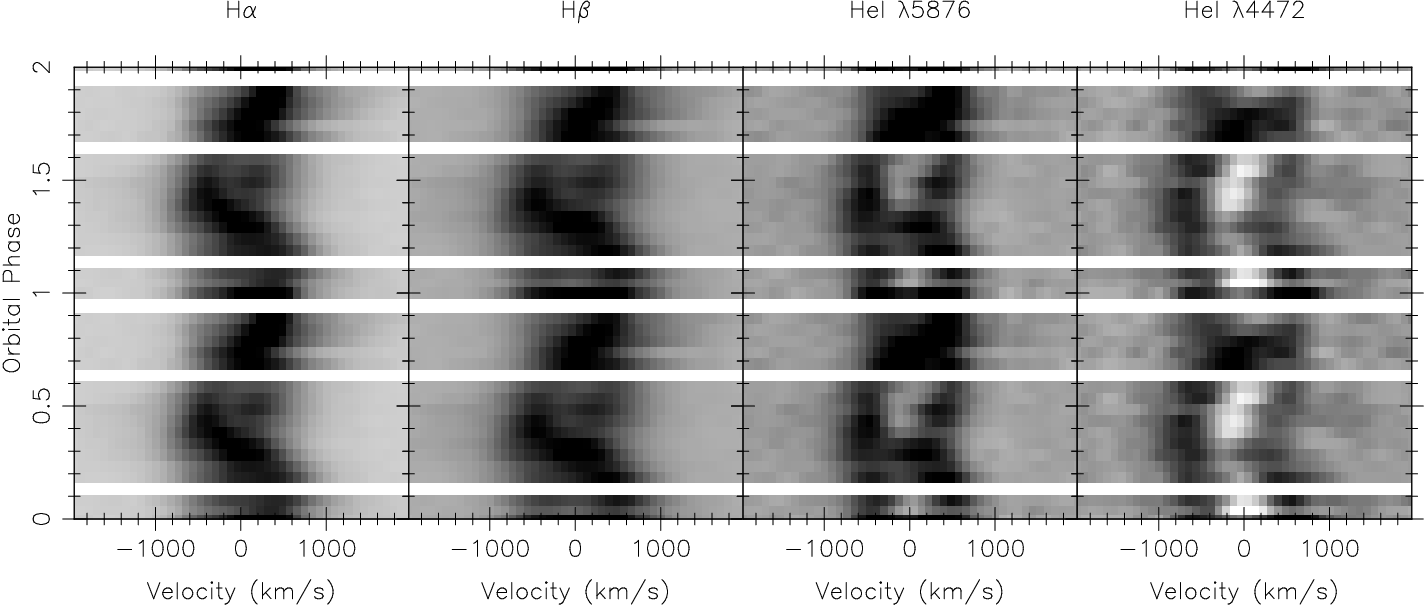}\\
  \vspace{0.5cm}
  \includegraphics[width=6cm,angle=-90]{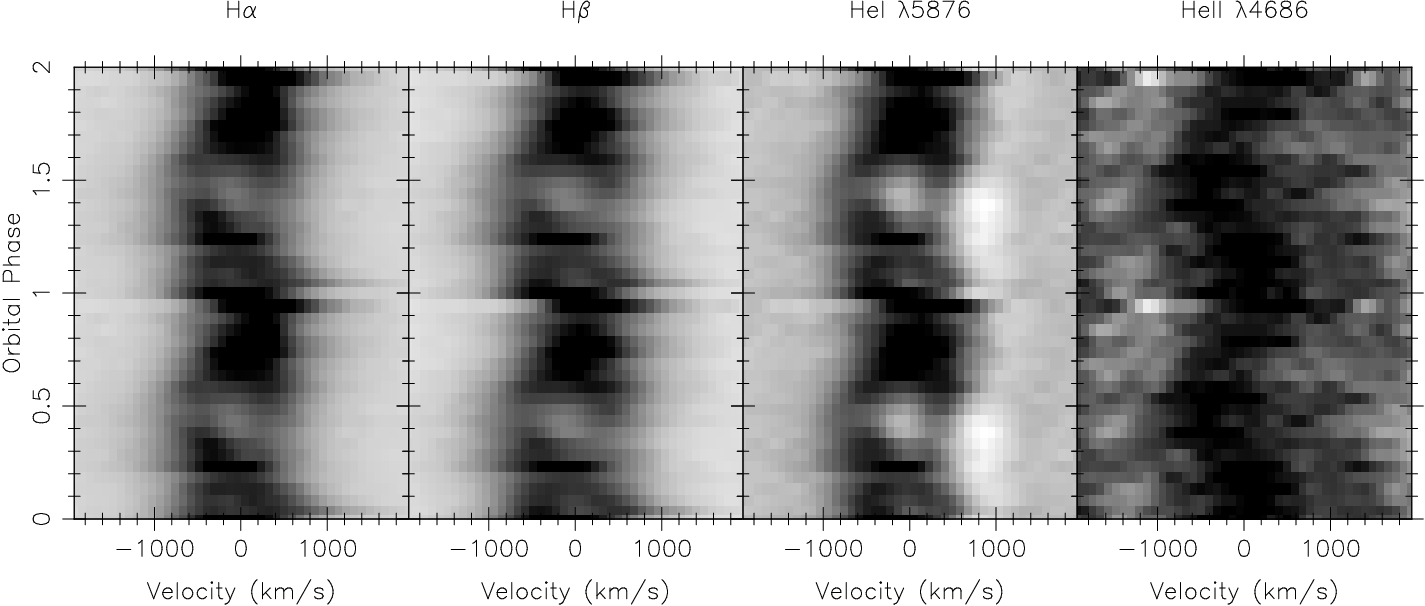}\\
  \vspace{0.5cm}
  \includegraphics[width=6cm,angle=-90]{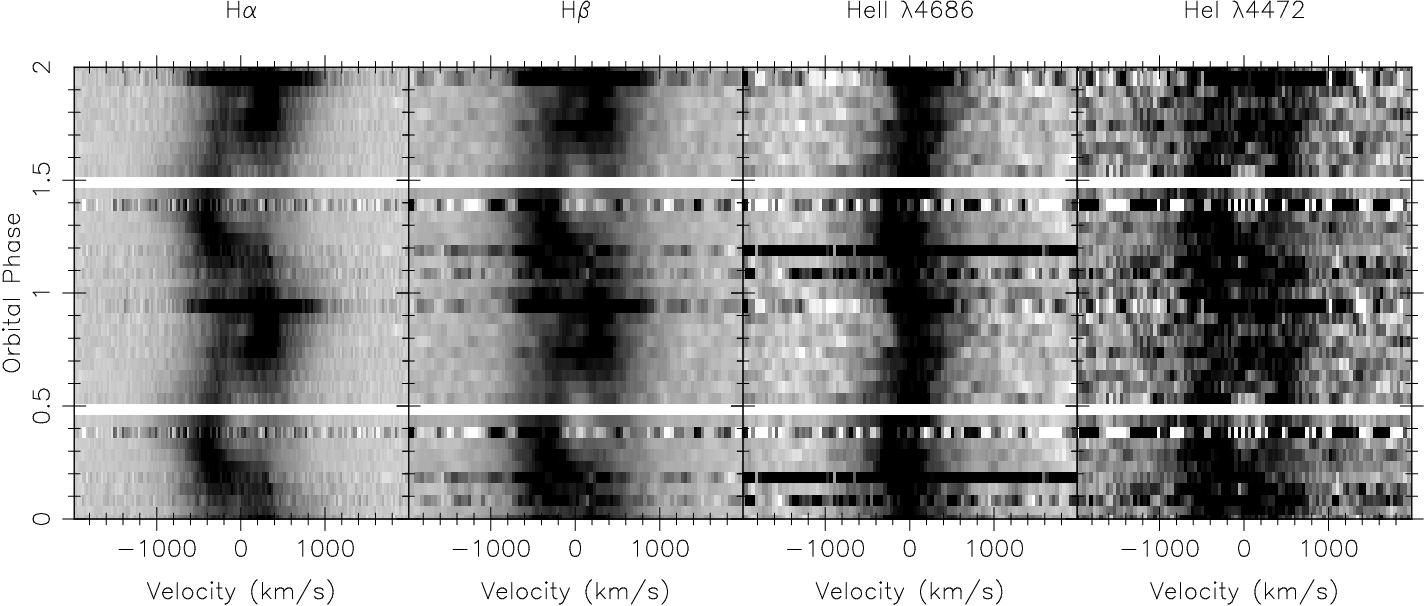}
\caption{\label{fig_trailed} Trailed spectra of (from \textit{top} to
\textit{bottom}) HS\,0129+2933, HS\,0220+0603, and HS\,0455+8513. The
individual spectra were averaged into 20 phase bins. Black represents
emission. Unsampled phase bins are indicated by a blank
spectrum. A full orbital cycle has been repeated for continuity.}
\end{figure*}

Trailed spectrograms of several emission lines of HS\,0129+2933,
HS\,0220+0603, and HS\,0455+8315 were constructed after re-binning the
spectra on to an uniform velocity scale centred on the rest wavelength
of each line. They are shown in Fig.~\ref{fig_trailed}. The \Ha~and
\Hb~line emission is dominated by a high-velocity emission component
which neither follows the phasing of the primary nor the
secondary. These S-waves reach their bluest velocity at $\varphi \sim
0.5$ in the three eclipsing systems and have a velocity amplitude
of $\ga 600~\kms$. Weaker emission can also be seen underneath the
dominant S-wave, possibly originating in the accretion disc. An
absorption component is observed moving across the lines from red to
blue, reaching maximum strength also at $\varphi \sim 0.5$.

The \Ion{He}{i} lines also display this high-velocity S-wave with the
same phasing as the Balmer lines, as well as the absorption component. The \Line{He}{i}{4472} line in HS\,0220+0603 shows
it for approximately three quarters of an orbit, going well below the
continuum.

Only HS\,0220+0603 and HS\,0455+8315 have a \Line{He}{ii}{4686} line
strong enough to produce a clear trailed spectrogram. While in the
former the emission seems to be entirely dominated by the
high-velocity component, two components are observed in the latter: a
low-velocity one (the strongest), and a weaker component which is
probably the same S-wave we observe in the Balmer and \Ion{He}{i}
lines. No significant absorption is detected in \Line{He}{ii}{4686}.

\begin{figure}
\centering
  \includegraphics[width=6cm,angle=-90]{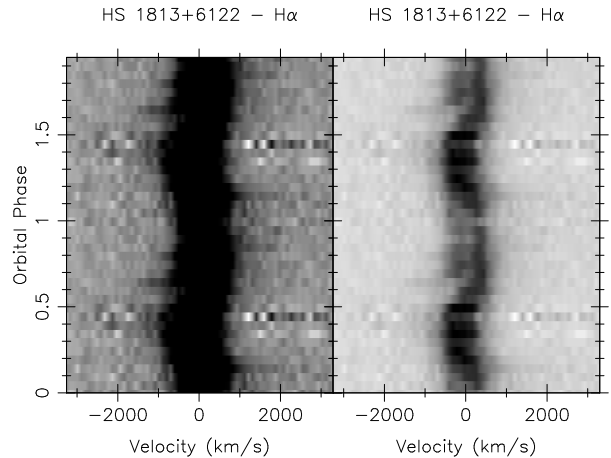}
\caption{\label{fig_hs1813hatrailed} \Ha~trailed spectra of
HS\,1813+6122. Black represents emission. The grey scale has been adjusted to enhance the high-velocity S-wave (\textit{left}) and the line core (\textit{right}). A whole cycle has been repeated.}
\end{figure}

In Fig.~\ref{fig_hs1813hatrailed} we present the \Ha~trailed spectrum
of HS\,1813+6122 after averaging the individual spectra into 20 orbital phase
bins. The line shows a high-velocity S-wave with maximum blue velocity
at $\sim -2000$\,\kms.

\subsection{The FUV spectrum of HS\,0455+8315}\label{sect_hst_hs0455}

The ultraviolet spectrum of HS\,0455+8315 (obtained at $\varphi \simeq 0.75$) displays very strong emission lines of He, C, N, O, and Si, with line flux ratios compatible with those observed in the majority of CVs
(Mauche, Lee \& Kallman \citeyear{maucheetal97-1}), indicating normal chemical abundances of the
donor star. The slope of the FUV continuum is nearly flat as observed in a number of deeply eclipsing SW\,Sex stars (e.g. DW\,UMa, \citealt{szkody87-2,kniggeetal04-1}; PX\,And,
\citealt{thorstensenetal91-1}; BH\,Lyn, \citealt{hoard+szkody97-1}), . It has been argued that a relatively cold structure shields the inner disc and the white dwarf from view in
high-inclination SW\,Sex stars during the high state, specifically
supported by the FUV detection of the hot white dwarf in DW\,UMa
during a low state \citep{kniggeetal00-1,araujo-betancoretal03-1}.

\section{SW\,Sex membership}

The five new CVs presented in this paper have very much in common. In their average spectra the Balmer and \Ion{He}{i} lines display both single- or double-peaked profiles. The double-peaked profiles are likely a consequence of phase-dependent absorption components as the trailed spectra show. The lines are also characterised by highly asymmetric profiles with enhanced wings. The trailed spectra reveal the presence of a
high-velocity emission S-wave in all the systems with extended wings reaching a maximum velocity between $\sim \pm 2000$ \kms~(HS\,1813+6122) and $\sim \pm
1000$ \kms~in the eclipsing systems. The tendency of non-eclipsing SW Sex stars to show broader S-waves may be evidence of emitting material with a vertical velocity gradient such as a mass outflow.

The radial velocity curves also show a distinctive SW Sex feature. They are
delayed with respect to the motion of the white dwarf, so that the
red-to-blue crossing takes place at $\varphi \sim 0.2$ instead of
$\varphi=0$. On the other hand, the eclipsing systems also display a
discontinuity around mid-eclipse, probably a rotational disturbance,
which indicates that part of the line emission comes from the
accretion disc.

All the features described above are defining characteristics of the
SW Sex stars (see \citealt{thorstensenetal91-1} and Rodr\'\i guez-Gil, Schmidtobreick \& G\"ansicke \citeyear{rodriguez-giletal07-1}). Even though each
system exhibits its own peculiarities (e.g. unusual spectral lines in
HS\,0220+0603 and strong \Line{He}{ii}{4686} emission in
HS\,0455+8315), they all share the characteristic SW\,Sex behaviour. We therefore classify them all as new SW\,Sex stars.

\section{The SW Sex stars in the context of CV evolution}

\subsection{How are the SW Sex stars discovered as CVs?}

CVs are found by a number of means. Many of them display abrupt
brightenings as a result of a dwarf nova eruption or a nova
explosion. They also show a rich variety of photometric variations
like eclipses, orbital modulations, rapid oscillations, ellipsoidal modulations, etc. Others have been discovered by their blue colour, the emission of X-rays, or the presence of strong emission lines in their optical spectra. 

The total number of definite members of the SW Sex class so far known
amounts to 35, out of which a remarkable number of 18 (51 per cent) has been
found in UV-excess surveys (i.e. blue colour). This is not surprising,
as the optical spectra of the SW\,Sex stars in the high-state is
characterised by a very blue continuum. On the other hand, 11 SW\,Sex stars (31 per cent) have been identified as
CVs from their emission line spectra, five of which were discovered in
the HQS. Only four and two SW\,Sex stars have been found because of their
brightness variability (including 3 novae) and X-ray emission,
respectively.

Any sort of CV search has its own selection effects, and the
classification of a CV as a SW Sex star is no exception. In fact, the
deep eclipses that many of the SW Sex stars show initially made the sample
clearly biased towards high inclination systems. This led many
authors to link the SW Sex phenomenon to a mere inclination effect. At
the last count, 13 out of a total of 35 SW Sex stars (37 per cent) do not
display eclipses and are \textit{bona fide} members of the
class. Although an inclination effect may certainly be important (see the case of HL Aqr in \citealt{rodriguez-giletal07-1}), the increasing number of non-eclipsing systems poses serious difficulties to any model resting solely on a high orbital inclination.

In the following section we discuss on the impact of SW Sex stars in the
(spectroscopic) HQS sample, which is unaffected by the high-inclination
selection effect.

\subsection{The role of the SW Sex stars in the big family of nova-like CVs}

\begin{figure}
\centering \includegraphics[width=\columnwidth,angle=0]{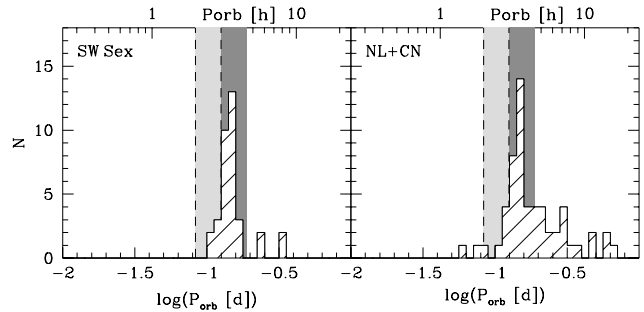}
\caption{\label{fig_swsex_porb} Period distributions of confirmed
  SW\,Sex stars (left panel) and nova-like variables and classical
  novae that are not known to exhibit SW\,Sex behaviour. The orbital period gap and the 3--4 h orbital period range are indicated in light and dark grey, respectively.}
\end{figure}

\begin{table}
\caption[]{\label{tab_swsex} The SW Sex stars.}

\begin{center}
\begin{tabular}{lccl}
\hline\noalign{\smallskip}
Object & $P_\mathrm{orb}$ (h) & Eclipses   & References\\
\hline\noalign{\smallskip}
V348\,Pup                  &  2.44 & Yes     & 1          \\
V795\,Her                  &  2.60 & No      & 2          \\
RX\,J1643.7+3402           &  2.89 & No      & 3          \\
V442\,Oph                  &  2.98 & No      & 4          \\
AH\,Men                    &  3.05 & No      & 5          \\ 
HS\,0728+6738              &  3.21 & Yes     & 6          \\  
HS\,0805+3822              &  3.21 & Grazing & 7,         \\
                           &       &         & this paper \\
SW\,Sex                    &  3.24 & Yes     & 8          \\
HL\,Aqr                    &  3.25 & No      & 5          \\
DW\,UMa                    &  3.28 & Yes     & 8          \\
SDSS\,J132723.39+652854.2  &  3.28 & Yes     & 9          \\
WX\,Ari                    &  3.34 & Grazing & 10	  \\	      
HS\,0129+2933              &  3.35 & Yes     & This paper \\ 
V1315\,Aql                 &  3.35 & Yes     & 8          \\
BO\,Cet                    &  3.36 & No      & 5          \\  
AH\,Pic                    &  3.41 & No      & 5          \\
VZ\,Scl                    &  3.47 & Yes     & 11         \\  
LN\,UMa                    &  3.47 & No      & 5          \\
RR\,Pic                    &  3.48 & Grazing & 12         \\
PX\,And                    &  3.51 & Yes     & 8          \\
V533\,Her                  &  3.53 & No      & 13         \\
HS\,1813+6122              &  3.54 & No      & This paper \\  
HS\,0455+8315              &  3.57 & Yes     & This paper \\ 
HS\,0220+0603              &  3.58 & Yes     & This paper \\
HS\,0357+0614              &  3.59 & No      & 14         \\
V380\,Oph                  &  3.70 & No      & 5          \\
BH\,Lyn                    &  3.74 & Yes     & 15  	  \\
UU\,Aqr                    &  3.93 & Yes     & 16         \\
LX\,Ser                    &  3.95 & Yes     & 17         \\
V1776\,Cyg                 &  3.95 & Yes     & 18         \\ 
LS\,Peg                    &  4.19 & No      & 19         \\
V347\,Pup                  &  5.57 & Yes     & 20         \\
RW\,Tri                    &  5.57 & Yes     & 21         \\ 
V363\,Aur                  &  7.71 & Yes     & 22         \\  
BT\,Mon                    &  8.01 & Yes     & 23         \\  
\noalign{\smallskip}\hline
\end{tabular}
\end{center}
\rmfamily {\it References}: 1 \cite{rodriguez-giletal01-2}; 2
\cite{casaresetal96-1}; 3 \cite{mickaelianetal02-1}; 4
\cite{hoardetal00-1}; 5 \cite{rodriguez-giletal07-1}; 6 \cite{rodriguez-giletal04-2};
7 \cite{szkodyetal03-2mn}; 8 \cite{thorstensenetal91-1}; 9 \cite{wolfeetal03-1}; 10 \cite{beuermannetal92-1}; 11 \cite{moustakas+schlegel99-1}; 12 Schmidtobreick, Tappert \& Saviane (\citeyear{schmidtobreicketal03-1}); 13 \cite{rodriguez-gil+martinez-pais02-1}; 14 \cite{szkodyetal01-1}; 15 Thorstensen, Davis \& Ringwald (\citeyear{thorstensenetal91-2}); 16
\cite{hoardetal98-1}; 17 Young, Schneider \& Shectman (\citeyear{youngetal81-2}); 18
\cite{garnavichetal90-1}; 19 \cite{tayloretal99-1}; 20
\cite{thoroughgoodetal05-1}; 21 Groot, Rutten \& van Paradijs (\citeyear{grootetal04-1}); 22
\cite{thoroughgoodetal04-1}; 23 Smith, Dhillon \& Marsh (\citeyear{smithetal98-1})
\end{table}

In Table~\ref{tab_swsex} we list the 35 known SW Sex stars along with
their orbital periods. Before doing any statistics we want to stress the fact that Table~\ref{tab_swsex} does not intend to present a definitive census of the SW Sex stars. This is because their defining characteristics are continuously evolving as we dig deeper into the understanding of this class of CVs. For comparison (and probably completeness) we also point the reader to Don Hoard's {\em Big List of SW Sex stars}.\footnote{http://spider.ipac.caltech.edu/staff/hoard/biglist.html}

Table~\ref{tab_swsex} shows that a significant 37 per cent of the family does not show eclipses, confirming that the high inclination requirement is merely a
selection effect. The orbital period distribution of the known sample
of SW Sex stars is presented in the left panel of
Fig.~\ref{fig_swsex_porb}. The combined distributions for non-SW Sex nova-likes
and nova remnants are also plotted for comparison. The nova-likes and
novae were selected from the Ritter \citep{ritter+kolb03-1}, CVcat
\citep{kubeetal03-1}, and Downes \citep{downesetal05-1}
catalogues. Only systems with a robust orbital period determination
are included.

These orbital period distributions reveal important features. About
40 per cent (35 out of 93) of the whole nova-like/classical nova
population (which is preferentially found above the period gap; only 12 are below it) are indeed SW Sex stars. Even more remarkable is
the fact that the SW Sex stars represent almost \textit{half} (26 out of
53) the CVs in the narrow $3-4.5$\,h orbital period range. Above
4.5\,h things radically change and only 14 per cent (4 out of 28) of the
nova-like/nova population are known to be SW\,Sex stars. The impact of SW Sex
stars in the gap is also striking, as 55 per cent of all the nova-like/nova
gap inhabitants are members of the class.

The non-SW Sex nova-like/nova $P_\mathrm{orb}$ distribution also shows
a significant fraction of systems in the $3-4.5$\,h interval. This nicely depicts the tendency of nova-likes to accumulate in the $3-4.5$\,h period
range. Therefore, it is very likely that more SW Sex stars are still to be found. In fact, time-resolved spectroscopic studies like the one reported in
\cite{rodriguez-giletal07-1} are revealing the SW\,Sex nature of many previously poorly studied nova-likes in this range. If the rate of detection and identification of SW Sex systems remains high the dominance of this class at the upper edge of the gap will eventually become even more pronounced.

It is possible, however, that these numbers are the result of a
selection effect as the majority of SW Sex stars are bright, making them
easily accesible to observations. Therefore, one can argue that a
proper characterisation of the whole population of CVs above the gap
needs to be made before addressing any conclusion. However, the fact the majority of SW Sex stars have orbital periods between 3 and 4.5 h is a well established fact.

\subsubsection{The impact of SW Sex stars on the HQS CV sample}

During the course of our spectroscopic search we have discovered 53
new CVs in the area/magnitude range covered by the HQS ($\simeq 13600$ square degrees/$17.5 \la B \la 18.5$; see \citealt{hagenetal95-1}). So far, we
have determined the orbital period for 43 of them in a huge
observational effort. Our preliminary results support the SW Sex
excess within the $3-4.5$\,h range. Remarkably, 54 per cent (7 out of 13) of
all the newly-discovered HQS nova-likes are indeed SW Sex stars, which
is in agreement with the distribution discussed above. This gives
further strength to the significancy of the observed (and still
unexplained) pile-up of SW Sex stars in the $3-4.5$\,h region.

\subsection{CV evolution and the SW Sex stars}
This accumulation of systems just above the period gap seriously
challenges our current understanding of CV evolution. The SW Sex stars
are intrinsically very luminous, as the brightness of systems like DW
UMa indicates. DW UMa is a SW Sex star likely located at a distance between $\sim 590-930$ pc \citep{araujo-betancoretal03-1} that has an average magnitude
of $V \sim 14.5$, even though it is viewed at an orbital inclination of $\sim
82$\degr. Therefore, in order to show such high luminosities, either
the SW Sex stars have an average mass transfer rate well above that of
their CV cousins, or another source of luminosity
exists. Neither the Rappaport, Joss \& Verbunt (\citeyear{rappaportetal83-1}) nor the (empirical)
Sills, Pinsonneault \& Terndrup (\citeyear{sillsetal00-1}) angular momentum loss prescriptions account for a
largely enhanced mass transfer rate ($\dot{M}$) in the period interval
where most of the SW Sex stars reside. In this regard, nuclear
burning has been suggested as an extra luminosity source
\citep{honeycutt01-1}, but the necessary conditions for the burning to
occur can only be found in the base of a magnetic accretion funnel,
suggesting a magnetic nature \citep{honeycutt+kafka04-1}. Temporary
cessation of nuclear burning would in principle explain the VY Scl low
states that many SW Sex and other nova-likes undergo. If this is true,
the majority of CVs above the period gap have to be magnetic, in stark contrast with a non-magnetic majority below the gap. However, the
fact that some dwarf novae (like HT Cas and RX And) also show low
states (e.g. \citealt{robertson+honeycutt96-1}; Schreiber, G\"ansicke \& Mattei \citeyear{schreiberetal02-1}) argues against this possibility, suggesting that the VY Scl states are likely the product of a decrease in the mass transfer rate from the donor star caused by starspots, as already proposed by
\cite{livio+pringle94-1}. The observational results of
\cite{honeycutt+kafka04-1} appear to support this hypothesis
\citep[see also e.g.][]{howelletal00-1}.

All of the above arguments point to accretion at a very high $\dot{M}$ as the most likely cause for the high luminosity observed in the SW Sex stars. One possibility could be enhanced mass transfer due to
irradiation of the inner face of the secondary star by a very hot
white dwarf. In fact, a number of nova-like CVs in the $\sim 3-4$ h
orbital period range (including the SW Sex star DW UMa) have been
observed to harbour the hottest white dwarfs found in any CV
\citep[see][]{araujo-betancoretal05-2}, with effective temperatures
peaking at $\sim 50\,000$ K. These high temperatures are most
likely the result of accretion heating, as CVs are thought to spend on
average $\sim 2$ Gyr as detached binary systems
\citep{schreiber+gaensicke03-1}, enough time for the white dwarfs to
cool down to $\la 8000$ K. Hence, the high effective temperatures
measured in the $3-4$ h range is a measure of a very high secular
mass accretion rate of $\sim5\times10^{-9}~\mathrm{M_\odot}\,\mathrm{yr}^{-1}$
\citep{townsley+bildsten03-1}, higher than predicted by angular
momentum loss due to magnetic braking. Irradiation of the donor star has been observed in the non-eclipsing SW Sex stars HL Aqr, BO Cet, and AH Pic \citep{rodriguez-giletal07-1}, which supports the above hypothesis. Alas, the question of why the SW Sex stars have the highest mass transfer rates is still
lacking a satisfactory explanation in the context of the current CV
evolution theory.

\subsection{A rich phenomenology to explore}

It is becoming apparent that the SW Sex phenomenon is not restricted to the (mainly) spectroscopic properties initially introduced by \citealt{thorstensenetal91-1}. These maverick CVs are now known to show a much more intricate behaviour. Therefore, only a comprehensive study of this rich phenomenology will definitely lead to a full understanding of the SW Sex stars.
In what follows we will review the range of common features exhibited by the SW\,Sex stars and which implications can be derived from them.

\subsubsection{Superhumps}

A significant one third of the SW\,Sex stars is known to show apsidal (positive) or/and nodal (negative) superhumps, which are large-amplitude photometric waves modulated at a period slightly longer or slightly shorter than the orbital period, respectively. Positive superhumps are believed to be the effect of an eccentric disc which is forced to progradely precess by the tidal perturbation of the donor star \citep{whitehurst88-1}. On the other hand, negative superhumps are likely linked to the retrograde precession of a warped accretion disc \citep{murrayetal02-1}. Vertical changes in the structure of the disc may be triggered by the torque exerted on the disc by the tilted, dipolar magnetic field of the secondary star. Apparently, positive and negative superhumps are independent and can either coexist or alternate with a time scale of years. 

The detection of superhumps in the SW\,Sex stars is of great importance as they exhibit the largest superhump period excesses. Therefore, they are fundamental in calibrating the period excess--mass ratio relationship for CVs \citep[see][]{pattersonetal05-3}.    

\subsubsection{Variable eclipse depth}  

The continuum light curves of some eclipsing members of the class reveal that the eclipse depth varies with a time-scale of several days. So far this variability has only been studied in PX\,And \citep{stanishevetal02-1,boffinetal03-1} and DW\,UMa \citep[][ not to be confused with the changing eclipse depth in different brightness states]{biro00-1,stanishevetal04-1}. In PX And, \citealt{stanishevetal02-1} identify this long periodicity with the precession period of the accretion disc, which may also be true for DW UMa. The actual mechanism is not yet understood, although the eclipses of a retrogradely precessing, warped disc may account for what is observed in PX\,And \citep{boffinetal03-1}.    

\subsubsection{Low states}

Half of the nova-likes known to undergo VY\,Scl faint states are SW\,Sex stars. During these events the system brightness can drop by up to $\sim 4-5$\,mag and can remain that low for months. As the disc warping effect mentioned above, these low states may be controlled by the strong magnetic activity of the donor star, and are believed to be driven by a sudden drop of the accretion rate from the secondary star due to large starspots located in the area around the inner Lagrangian point $\mathrm{L}_1$ \citep{livio+pringle94-1}.

Interestingly, the disced VY\,Scl stars concentrate in the $3-4.5$\,h orbital period stripe as the SW\,Sex stars do. It would therefore not be a surprise to find that all nova-likes in that range are actually VY\,Scl stars and even SW\,Sex stars. Nevertheless, the presence of large starspots around the $\mathrm{L}_1$ point still have to be observationally confirmed. Using Roche tomography techniques, Watson, Dhillon \& Shahbaz (\citeyear{watsonetal06-1}) discovered a heavily spotted secondary star in AE\,Aqr ($P_\mathrm{orb}=9.88$\,h) with a spot distribution resembling that of other rapidly rotating, low-mass field stars. A stellar atmosphere plagued with spots has been also observed in the pre--CV V471\,Tau \citep[$P_\mathrm{orb}=12.51$\,h,][]{hussainetal06-1}. Unfortunately, the high spectral and time resolution required to image the M-dwarf donor in a CV with $P_\mathrm{orb} \simeq 3.5$\,h at a distance of many hundred parsecs is currently beyond observational reach.  

\subsubsection{Quasi-periodic oscillations}

On the grounds of short-term variability, the SW\,Sex stars are also characterised for exhibiting quasi-coherent modulations in their light curves. In a compilation of CVs displaying rapid oscillations, \cite{warner04-1} lists 9 (only two deeply eclipsing) SW\,Sex stars known to have quasi-periodic oscillations (QPOs) with a predominant time-scale of $\sim 1000-2000$\,s. For example, the non-eclipsing SW\,Sex stars V442\,Oph and RX\,J1643.7+3402 show strong QPO signals dominating over other underlying higher-coherence oscillations \citep{pattersonetal02-1}. These results are based on hundreds of
hours of photometric data, whose power spectra showed rapid frequency
changes of the main signals in less than a day. Similar rapid variability was also detected in the optical light curve of the SW\,Sex star HS\,0728+6738 \citep{rodriguez-giletal04-2}, with a prominent signal (coherent for at least 20 cycles) at $\sim 600$\,s.

\subsubsection{Emission-line flaring}

In connection with the QPO activity, the fluxes and EWs of the emission lines of some SW\,Sex stars show modulations at the same time-scales (a phenomenon known as emission-line flaring): $\sim 1800$\,s in BT\,Mon \citep{smithetal98-1}, $\sim 2000$\,s in LS\,Peg \citep{rodriguez-giletal01-1}, $\sim 1400$\,s in V533\,Her \citep{rodriguez-gil+martinez-pais02-1}, $\sim 1800$\,s in DW\,UMa (V. Dhillon, private communication), $\sim 2400$\,s in RX\,J1643.7+3402 (Mart\'\i nez-Pais, de la Cruz Rodr\'\i guez \& Rodr\'\i guez-Gil \citeyear{martinez-paisetal07-1}), and $\sim 1200$\,s in BO\,Cet \citep{rodriguez-giletal07-1}. Remarkably, the radial velocities measured in the last two objects are also modulated at the flux/EW periodicities, which suggests that emission-line flaring has to do with the dynamics of the line emitting source, and is not due to e.g. random fluctuations in the disc continuum emission. On the other hand, although DW UMa does exhibit emission-line flaring in the optical, such line variability was not detected in the far ultraviolet \citep{hoardetal03-1}. Since similar flaring in the optical is observed in the intermediate polar CVs \citep[IPs; e.g. FO\,Aqr][]{marsh+duck96-2} caused by the rotation of the magnetic white dwarf, all the described rapid variations seen in many SW\,Sex stars have been associated to the presence of magnetic white dwarfs \citep{rodriguez-giletal01-1,pattersonetal02-1}. Nevertheless, the far ultraviolet data of DW UMa presented by \cite{hoardetal03-1} can also be explained with a stream overflow model, but do not necessary exclude a magnetic scenario either.

\subsubsection{Variable circular polarisation}

Despite the fact that circular polarisation is not commonly detected in the majority of IPs, it is a \textit{sine qua non} condition for IP membership \citep[see further requirements in][]{patterson94-1}. The cyclotron radiation emitted by the accretion columns (built up by disc plasma forced by the magnetic field to supersonically fall on to the white dwarf surface) is known to be circularly polarised, thus the detection of a significant level of circular polarisation is a unequivocal sign of magnetic accretion. This, and the possible magnetic nature of the QPO and line flaring activity prompted to the search for circular polarisation in the SW\,Sex stars.

\cite{rodriguez-giletal01-1} found circular polarisation modulated at 1776\,s with a peak-to-peak amplitude of 0.3 per cent in LS\,Peg. Remarkably, the flaring observed in the \Hb~high-velocity S-wave was modulated at 2010\,s, which is just the synodic period between the polarisation period and the orbital period. V795\,Her also revealed variable circular polarisation with a periodicity of 1170\,s (or twice that) and showed an increasing polarisation level with wavelength \citep{rodriguez-giletal02-1} as is expected for cyclotron emission. In addition, RX\,J1643.7+3402 shows circular polarisation modulated at 1163\,s \citep{martinez-paisetal07-1}. 

The 1776-s period of LS\,Peg was confirmed by Baskill, Wheatley \& Osborne (\citeyear{baskilletal05-1}) who detected a coherent modulation at 1854\,s in {\sl ASCA} X-ray light curves. The coincidence of both periods indicates a common origin, with the X-ray period reported by \citeauthor{baskilletal05-1} likely being a more accurate measurement of the white dwarf spin period.

Although polarimetric studies of many other SW\,Sex stars have to be done, the results obtained so far suggest that magnetic accretion may play an important role in the SW\,Sex phenomenon. However, with the few such studies so far at hand it is not possible to address any conclusion regarding the impact of magnetism in the whole class.

~~

Despite the broad implications in our understanding of accretion that
the study of the SW\,Sex stars have, the quest for a successful, global
explanation of the phenomenon has been unfruitful so far. This, in addition to the (yet unexplained) fact that the majority of SW\,Sex stars (and many nova-likes) largely populate the narrow orbital period stripe between 3 and 4.5 hours, are seriously shaking the grounds on which CV evolution theory stands. Further study of these maverick systems will certainly provide fundamental clues to our understanding of CV evolution.

\section*{Acknowledgments}
To the memory of Emilios Harlaftis.
~\\
~\\
\indent AA thanks the Royal Thai Government for a
studentship. BTG was supported by a PPARC Advanced Fellowship,
respectively. MAPT is supported by NASA LTSA grant NAG-5-10889. RS is
supported by the Deutsches Zentrum f\"ur Luft und Raumfahrt (DLR) GmbH
under contract No.\,FKZ \mbox{50 OR 0404}.  AS is supported by the
Deutsche Forschungsgemeinschaft through grant Schw536/20-1. The HQS
was supported by the Deutsche Forschungsgemeinschaft through grants
Re\,353/11 and Re\,353/22.

This paper includes data taken at The McDonald Observatory of The University of Texas at Austin. It is also based in part on
observations obtained at the German-Spanish Astronomical Center, Calar
Alto, operated by the Max-Planck-Institut f\"{u}r Astronomie,
Heidelberg, jointly with the Spanish National Commission for
Astronomy;
on observations made at the 1.2m telescope, located at Kryoneri
Korinthias, and owned by the National Observatory of Athens, Greece; 
on observations made with the Isaac Newton Telescope, which is
operated on the island of La Palma by the Isaac Newton Group in the
Spanish Observatorio del Roque de los Muchachos of the Instituto de
Astrof\'\i sica de Canarias (IAC); 
on observations made with the 1.2m telescope at the Fred Lawrence Whipple Observatory, a facility of the Smithsonian Institution;
on observations made with the IAC80 telescope, operated on the island
of Tenerife by the IAC in the Spanish Observatorio del Teide;
on observations made with the OGS telescope, operated on the island of
Tenerife by the European Space Agency, in the Spanish Observatorio del
Teide of the IAC;
on observations made with the Nordic Optical Telescope, operated
on the island of La Palma jointly by Denmark, Finland, Iceland,
Norway, and Sweden, in the Spanish Observatorio del Roque de los
Muchachos of the IAC;
and on observations made with the NASA/ESA Hubble Space Telescope,
obtained at the Space Telescope Science Institute, which is operated
by the Association of Universities for Research in Astronomy, Inc.,
under NASA contract NAS 5-26555.

This publication makes use of data products from the Two Micron All Sky Survey, which is a joint project of the University of Massachusetts
and the Infrared Processing and Analysis Center/California Institute of
Technology, funded by the National Aeronautics and Space Administration
and the National Science Foundation.

\bibliographystyle{mn2e}
\bibliography{mn-jour,aabib}

\end{document}